\documentclass[twocolumn]{openjournal}

\usepackage{lipsum}

\usepackage{xcolor}
\usepackage{textgreek}
\usepackage[utf8]{inputenc}
\usepackage[english]{babel}

\usepackage{hyperref}
\hypersetup{
    unicode, 
    colorlinks=true,
    linkcolor=linkcolor,
    citecolor=linkcolor,
    filecolor=linkcolor,
    urlcolor=linkcolor,
}
\usepackage{color,colortbl}
\definecolor{linkcolor}{rgb}{0.0,0.3,0.5}
\usepackage{tensind}
\tensordelimiter{?}
\DeclareGraphicsExtensions{.bmp,.png,.jpg,.pdf}
\usepackage{verbatim}
\usepackage[normalem]{ulem}
\usepackage{orcidlink}
\usepackage{soul}

\usepackage{makecell}
\usepackage{subfigure}
\usepackage{float}
\usepackage{graphicx}	
\usepackage{amsmath}	
\usepackage{amssymb}	
\usepackage{newtxtext,newtxmath}
\usepackage{multirow}

\begin{document}
\title{Transient QPOs of \textit{Fermi}-LAT blazars with Linearly Multiplicative Oscillations}

\author{P. Pe\~nil$^{1,\ast}$\orcidlink{0000-0003-3741-9764},
        J. Otero-Santos$^{2,\ast}$\orcidlink{0000-0002-4241-5875},
        A. Circiello$^{1,\ast}$,
        A. Banerjee$^{1}$ \orcidlink{0000-0001-7796-8907}, 
        S. Buson$^{3,4}$\orcidlink{0000-0002-3308-324X},
        A. Rico$^{1}$\orcidlink{0000-0001-5233-7180},
        M. Ajello$^{1}$\orcidlink{0000-0002-6584-1703},
        S. Adhikari$^{1}$\orcidlink{0009-0006-1029-1026}
        }
\thanks{$^\ast$ppenil@clemson.edu, jorge.otero@pd.infn.it, acircie@clemson.edu}

\affiliation{$^1$Department of Physics and Astronomy, Clemson University, Kinard Lab of Physics, Clemson, SC 29634-0978, USA}
\affiliation{$^2$Istituto Nazionale di Fisica Nucleare, Sezione di Padova, 35131 Padova, Italy}
\affiliation{$^3$Julius-Maximilians-Universit\"at W\"urzburg, Fakultät f\"ur Physik und Astronomie, Emil-Fischer-Str. 31, D-97074 W\"urzburg, Germany}
\affiliation{$^4$Deutsches Elektronen-Synchrotron DESY, Platanenallee 6, 15738 Zeuthen, Germany}

\begin{abstract}
    We present a study on the detection and characterization of transient quasi-periodic oscillations (QPOs) in the $\gamma$-ray emission of blazars 4C +31.03, MG1 J123931+0443, and PKS 1622$-$253. Using light curves derived from \textit{Fermi} Large Area Telescope data, we investigate oscillatory patterns characterized by periodic multiplicative amplitudes that vary linearly over time. By segmenting the light curves into increasing and decreasing trends, we analyze each segment independently, allowing for precise measurements of both the periodicity and long-term variations. To interpret these QPOs, we explore various theoretical scenarios that could explain their origin and underlying physical mechanisms. The variability observed in 4C~+31.03 is more consistent with a stochastic process, whereas the periods estimated for MG1~J123931+0443 and PKS~1622$-$253 align with the precessional dynamics expected from binary supermassive black hole systems. However, the current results remain tentative and do not allow for a definitive conclusion.
\end{abstract}


\maketitle

\section{Introduction} \label{sec:intro}
Blazars are among the most intriguing and powerful types of extragalactic sources. These highly energetic objects are characterized by relativistic jets that are closely aligned with our line of sight \citep[e.g.,][]{wiita_lecture}, making them prominent sources of radiation across the electromagnetic spectrum, from radio waves to $\gamma$ rays \citep[e.g.,][]{urry_multiwavelengh, penil_2020, penil_mwl}. They can be divided into two subclasses, Flat Spectrum Radio Quasars (FSRQs) and BL Lacertae objects, depending on the presence or absence of features in their optical spectra \citep{urry_n}. One of the most significant aspects of blazars is their variability across multiple timescales. While this variability is often associated with stochastic phenomena \citep[][]{ruan_stocastic, covino_negation}, some blazars exhibit quasi-periodic oscillations (QPOs) in their emissions \citep[e.g.,][]{zhou_PKS_2247_131, sagar_pg1553}. In practical terms, a QPO is typically identified when at least three cycles are observed with peak separations consistent \citep[e.g.,][]{vaughan_criticism, penil_kink_2025}. These criteria allow for performing a study of the observed modulation that could reflect a physically meaningful variability rather than random stochastic fluctuations.

QPOs are of particular interest because they suggest the presence of coherent, cyclic processes within the blazar system, which can provide clues about the physical mechanisms governing their cores and relativistic jets. Potential explanations for QPOs include variations in the accretion rate \citep[e.g.,][]{gracias_modulation_disk, fichet_increase_trends_shock_2022} or dynamical interactions involving a secondary supermassive black hole (SMBH) within the blazar's core \citep[e.g.,][]{sandrinelli_redfit, sobacchi_binary}. Both scenarios can induce sustained, periodic variations in the observed emission. 

Recently, \citet[][]{penil_2025} and \citet[][]{alba_ssa} reported blazars exhibiting long-term trends in their $\gamma$-ray emission (an increase or decrease in flux over time). Additionally, \citet[][]{alba_ssa} found that some of these blazars also display periodic patterns superimposed on these trends. Two distinct types of oscillations associated with these long-term variations were observed: additive oscillations, where the amplitude of the flux oscillations remains constant and independent of changes in the long-term trend \citep[][]{penil_2022_multiwave}, and multiplicative oscillations, where the amplitude scales proportionally with variations in the trend. These findings suggest that periodic behavior in blazars can manifest in different ways, potentially providing insights into the physical mechanisms governing their variability.  

In this paper, we analyze the QPOs in the $\gamma$-ray light curves (LCs) of three blazars using data from the \textit{Fermi} Large Area Telescope \citep[LAT,][]{fermi_lat}. These QPOs are transient, as they persist for a finite duration, typically spanning only a few cycles over several years. Their limited duration, relative to the total observational time, suggests that they are not continuous features but rather episodic phenomena that appear and disappear over time. These QPOs exhibit a distinctive trend in their oscillatory behavior: the amplitude of the oscillations varies over time, showing a gradual increase followed by a decrease. Different theoretical models have been proposed to explain the presence of such amplitude trends. One possibility is the interaction of an SMBH binary system with the surrounding gas, which could lead to changes in the accretion rate, thus affecting the amplitude of the QPOs over time \citep[][]{sagar_pg1553}. Another scenario involves magnetic reconnection events triggered by instabilities in the jet \citep[][]{giannios_trend_magnetic}. By studying these blazars, we aim to expand our understanding of QPOs with multiplicative oscillations and their possible underlying physical origins.

This paper is organized as follows. In $\S$\ref{sec:sample}, we introduce the sources analyzed in this study and discuss the key aspects of the \textit{Fermi}-LAT data utilized. In $\S$\ref{sec:methodology}, we describe the methodology used to characterize these QPOs. $\S$\ref{sec:theoretical_model} provides an overview of the theoretical models considered to interpret the QPOs observed in our blazar sample. $\S$\ref{sec:flux_analysis} presents the results of the flux analysis for each blazar. Finally, $\S$\ref{sec:summary} summarizes our main findings and conclusions.

\begin{figure*}
	\centering
        \includegraphics[scale=0.223]{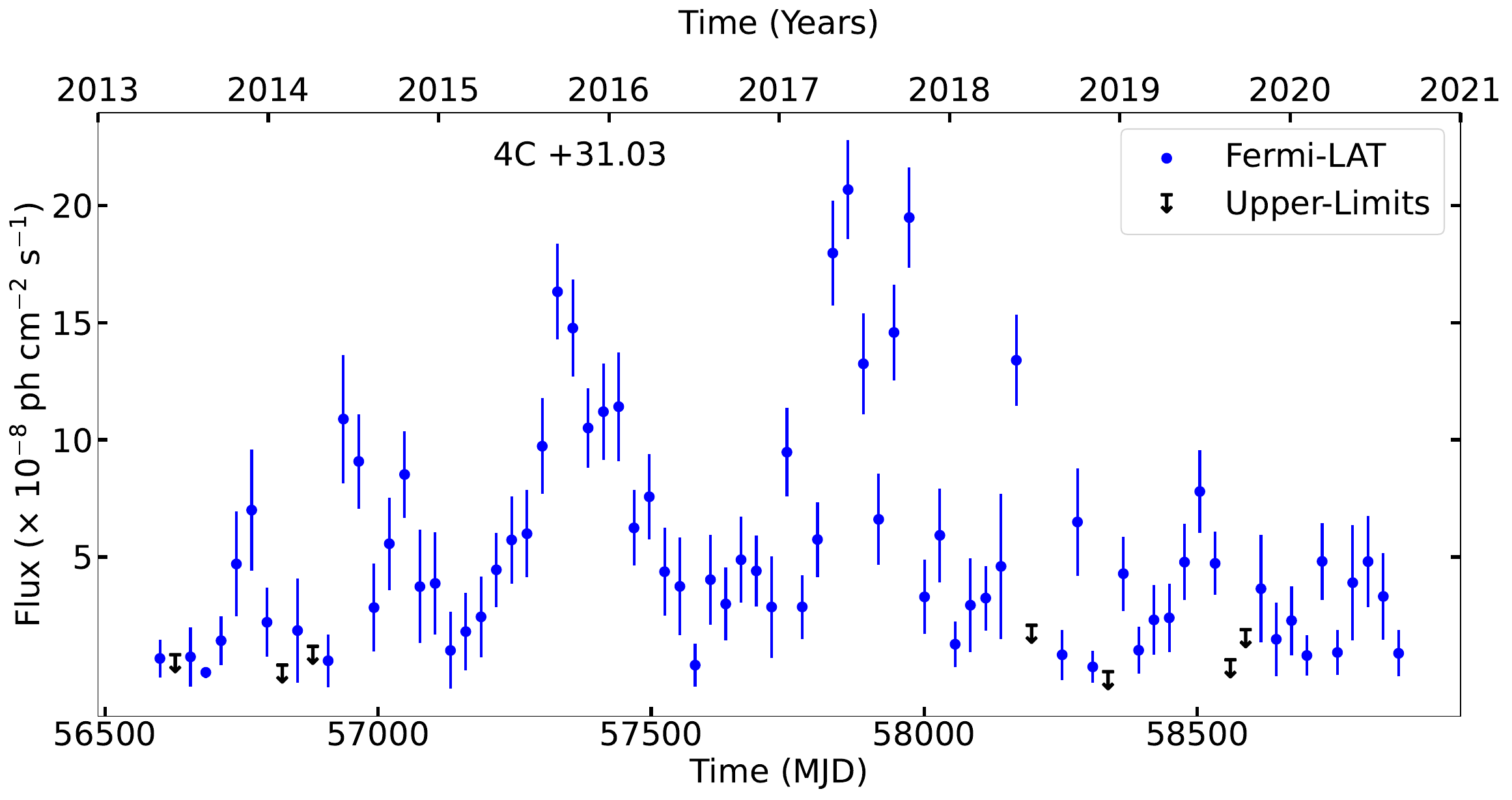}
	\includegraphics[scale=0.223]{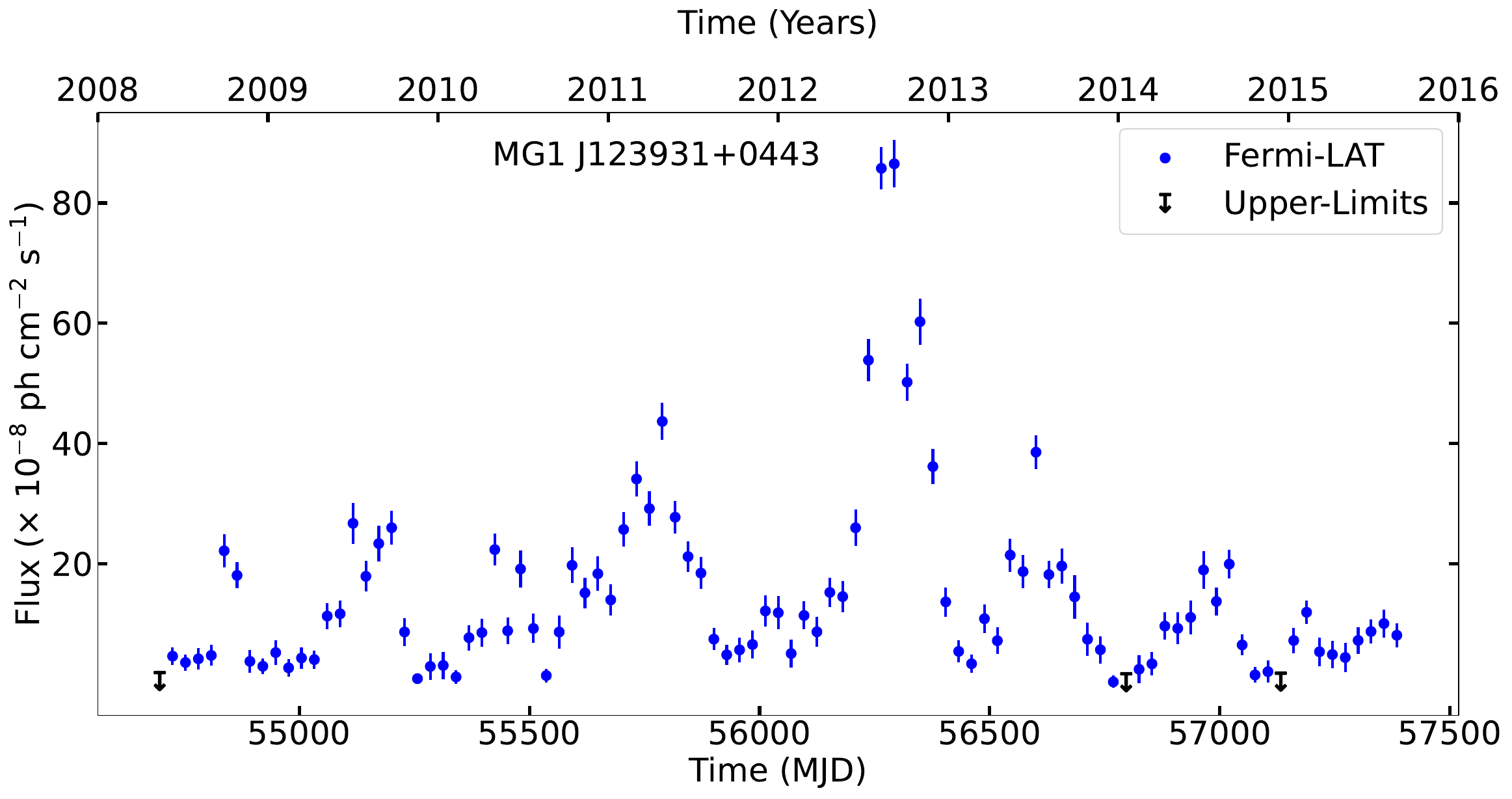}
        \includegraphics[scale=0.223]{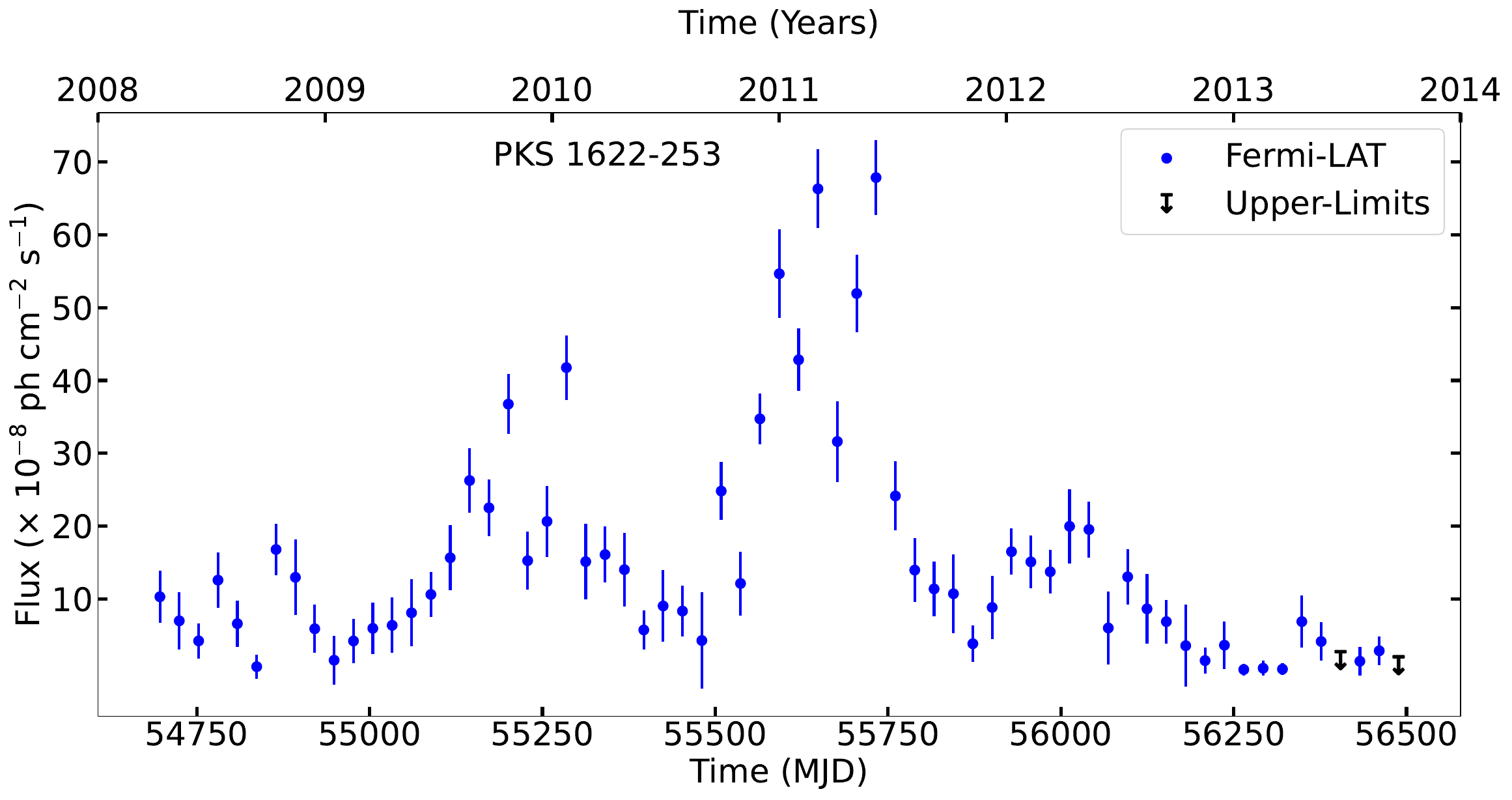}
	\caption{Segments of LCs of the blazars presented in Table \ref{tab:candidates_list} with signatures of transient QPOs. \textit{Top left}: 4C +31.03. \textit{Top right}: MG1 J123931+0443. \textit{Bottom}: PKS 1622$-$253. These segments are further analyzed in detail in $\S$\ref{sec:flux_analysis}.} \label{fig:qpos_lcs}
\end{figure*}

\section{Source Selection}\label{sec:sample}
We employed a sample of sources exhibiting long-term variability in their $\gamma$-ray emission. In particular, we analyzed 1,492 variable jetted AGN from the 4FGL-DR2 catalog \citep[][]{4fgl_dr2}. This sample was systematically examined for long-term trends in \citet[][]{penil_2025}. Additionally, a subsample of these $\gamma$-ray LCs was further investigated in a detailed periodicity study by \citep[][]{alba_ssa} using a novel method described in $\S$\ref{sec:methodology}. 

A systematic study was performed by applying a pipeline consisting of different processing stages \citep[][]{penil_2025_24, penil_2025_4fgl}. As a result, 998 objects with a period with no statistical significance ($\leq$1.5$\sigma$) were selected \citep[][]{penil_2025_4fgl}. Moreover, 337 objects were selected in the first analysis stage performed in \citet[][]{penil_2025} for showing some evidence of the presence of a long-term trend\footnote{This evidence comes from performing periodic analyses on both detrended and non-detrended LCs, allowing us to identify frequency-domain changes potentially introduced by long-term trends \citep[][]{penil_2025}.}, suggesting potential underlying processes that required further investigation \citep[][]{penil_2025}. 

A cross-check of both subsamples resulted in a new total of 93 coincident objects. Although the periods and trends identified in \citet[][]{penil_2025_4fgl, penil_2025} were statistically not significant, in this work, we revisit the previous subset as candidates of interest. These sources show potential indications of structured variability as periodic oscillations and long-term coherence that justify further examination beyond the initial automated classification.

These 93 objects are further analyzed with the Continuous Wavelet Transform \citep[CWT][]{wavelet_torrence}, a method that decomposes signals in both the time and frequency domains simultaneously. By applying the CWT, we can identify the presence of potential QPOs and locate the specific temporal segments of the LC where these QPOs occur. As a result, the CWT analysis of the sources reveals potential QPOs in specific LC regions, identifying them as transient QPO candidates. Through this analysis, we identify eleven blazars exhibiting transient QPO signatures. Among these, three blazars exhibited transient QPOs potentially explainable by the curved jet model \citep[e.g.,][]{sarkar_curved_jet, anuvab_curve_jet}. In this scenario, periodic variations arise from changes in the viewing angle, which modulate the observed flux through differential Doppler boosting as emitting regions travel along the curved geometry of the jet. Subsequent analysis confirmed that two of these blazars were consistent with this jet model \citep[][]{penil_curve_jet_2025}. In this paper, we refine our selection further by focusing specifically on three FSRQs, 4C +31.03, MG1 J123931+0443, and PKS 1622$-$253, which show oscillations combined with linear trends (Table~\ref{tab:candidates_list}). The analyzed segments of their LCs are presented in Figure \ref{fig:qpos_lcs}. The remaining five sources exhibit more complex QPO structures that cannot be adequately described by the linear-trend framework used in this study. Their analysis will require exploring alternative scenarios, such as quadratic long-term trends \citep[][]{quadratic_trend_binary_combi}, or exponential rise-decay oscillations \citep[][]{exponential_shape}.

The remaining objects of the sample present a more complex QPO structure, which will require the analysis of the application of alternative scenarios as quadratic trends \citep[][]{quadratic_trend_binary_combi}.  

For this analysis, we use the 28-day binned LCs above an energy of 0.1~GeV reported in \citet{penil_2025} and \citet[][]{alba_ssa}. The 28-day binning allows us to capture and analyze long-term variations in the $\gamma$-ray data while reducing the impact of short-term variability. Details of the Fermi-LAT data reduction procedure are provided in \citet{penil_2025_4fgl}.

\begin{table*}
\centering
\caption{List of blazars studied in this work, including their \textit{Fermi}-LAT source name, coordinates, AGN type, redshift, and association name. The sample consists of 3 flat-spectrum radio quasars (FSRQ). Additionally, we provide the specific LC segments exhibiting transient signatures, along with their corresponding dates in Modified Julian Date (MJD).}
\label{tab:candidates_list}

\resizebox{\textwidth}{!}{%
\begin{tabular}{ccccccc}
\hline\hline
4FGL Source Name & RA(J2000) [deg] & Dec(J2000) [deg] & Type & Redshift & Association Name & LC Segment [MJD] \\
\hline
J0112.8+3208 & 18.22 & 32.13 & FSRQ & 0.603 & 4C +31.03 & 56600--58900 \\
J1239.5+0443 & 189.88 & 4.72 & FSRQ & 1.761 & MG1 J123931+0443 & 54700--57400 \\
J1625.7$-$2527 & 246.44 & $-$25.46 & FSRQ & 0.786 & PKS 1622$-$253 & 54700--56500 \\
\hline
\end{tabular}%
}
\end{table*}

\section{Characterization of the QPOS} \label{sec:methodology}
In this section, we outline the methodological approaches used to analyze the LCs and characterize the observed QPOS in each blazar. 

\subsection {Methods for the Search of QPOs}
Two techniques are employed to identify QPOs in our blazar sample: the Singular Spectrum Analysis \citep[SSA,][]{ssa_greco, SSA_algorithm} and the Generalized Lomb-Scargle Periodogram  \citep[GLSP,][]{lomb_gen}.

SSA is a non-parametric method that decomposes a time series into its principal components, isolating trends, oscillating components, and noise. This decomposition allows for the identification of quasi-periodic components that may be hidden in the original noisy data \citep[][]{alba_ssa}. In QPO searches, SSA is advantageous because it does not require assumptions about the underlying periodicity, making it robust for detecting QPOs of varying frequencies and amplitudes.

Next, we apply the GLSP to this oscillatory component to determine the period and its associated significance, following the methodology outlined by \citet{alba_ssa}. The GLSP extends the traditional Lomb-Scargle Periodogram \citep[LSP,][]{Lomb_1976, Scargle_1982}, which is widely used for detecting periodic signals in red-noise-dominated and unevenly-spaced data. GLSP improves the classic method by incorporating the ability to handle non-sinusoidal periodicities and by weighting data points according to their uncertainties. However, we note that it can be affected by irregularities in the amplitudes of the oscillations, as shown by \cite{penil_flare_2025}. This two-step approach improves our ability to detect periodic signals in blazars by effectively separating the underlying oscillations from noise.

Finally, the uncertainties associated with the periods reported by the GLSP and SSA are estimated using the half width at half maximum of the corresponding peaks \citep[e.g.,][]{otero_mwl}. 

\subsection {Test Statistics of the QPOs}
QPO searches in time-series data are often limited by noise, which can mask real periodic signals. Red noise, with its erratic brightness variations and increased power at lower frequencies, is a common issue in blazar time series that can lead to false QPO detections \citep[][]{vaughan_criticism}. Additionally, Poisson noise caused by the random detection of photons adds a white noise component, particularly affecting blazar LCs in $\gamma$-rays. These combined noise sources make it challenging to reliably identify true QPOs in observational data.

To address the challenges posed by the noise, we simulate 150,000 artificial LCs following the approach of \citet{emma_lc}. This method creates synthetic LCs that mimic the power spectral density (PSD) and probability density function (PDF) of the original time series. The PSD is modeled by a power law, $A*f^{-\beta}+C$, where $A$ is the normalization, $\beta$ is the spectral index, $f$ is frequency, and $C$ represents Poisson noise. These parameters are estimated using Maximum Likelihood Estimation and Markov Chain Monte Carlo (MCMC) analysis\footnote{We utilize the \textsc{Python} package \texttt{emcee}.}. On the other hand, the PDF is described by a log-normal distribution.

\subsection {Results}
Using the previous methodology, we estimate the QPOs in the selected segments of our blazar sample. For 4C +31.03, the GLSP analysis suggests a primary peak at 525$\pm$60 days (2.7$\sigma$) and a secondary peak at 288$\pm$37 days (0.9$\sigma$), while for SSA, the primary peak is hinted at 516$\pm$88 days (0.7$\sigma$); however, it does not show any significant signature of the secondary peak at 288$\pm$37~days (0.1$\sigma$). Both methods report a similar period of $\approx$520 days but show notable differences in significance. Additionally, only the SSA hints at the presence of the secondary period at 288~days, with, however, a rather low significance. 

The higher confidence in SSA suggests it effectively reduces noise, enhancing the peak's significance. We also note that, as reported by \cite{penil_flare_2025}, amplitude changes in recurrent or periodic flares such as the ones observed here can have a very strong impact on the significance derived for such QPO patterns.

In the case of MG1 J123931+0443, the QPO estimates are 347$\pm$27 days (1.7$\sigma$) and 297$\pm$20 days (1.7$\sigma$) for SSA and GLSP, respectively. The difference in the period suggests that GLSP might be influenced by long-term trends, affecting period detection accuracy. By isolating trends, SSA provides a more precise estimate in such cases. However, their similar significance levels indicate the presence of a transient QPO in this source.

Finally, for PKS 1622$-$253, the results are 433$\pm$62 days (1.3$\sigma$) and 456$\pm$65 days (0.7$\sigma$) for SSA and GLSP, respectively. The periods are similar, but SSA shows a higher significance. However, the low significance in both cases suggests the periodicity remains tentative.

\section{Theoretical Models} \label{sec:theoretical_model}
QPOs observed in the $\gamma$-ray LCs of blazars have been extensively studied, as they provide valuable insights into the dynamic processes occurring in the cores of these objects. Recently, \citet[][]{penil_2025} reported that some blazars exhibit an associated long-term trend in their emission. These trends influence the amplitude of both aperiodic and periodic oscillations by modulating their multiplicative values over time as the long-term trend evolves. In some cases, QPOs appear superimposed on these long-term variations, suggesting a complex interplay between periodic oscillations and secular emission changes \citep[][]{alba_ssa}. This type of QPO suggests that they may result from multiple interdependent processes, indicating a complex system where independent phenomena could interact to generate QPOs and modulate their amplitudes multiplicatively. 

\subsection{Accretion Disk-Origin Scenarios}

The intrinsic temporary structure of the QPOs suggests that they could arise from a range of transient phenomena. For instance, accretion disk instabilities can lead to fluctuations in the mass accretion rate \citep[][]{abdo_variability}, which, when propagating inward, modulate the energy injection into the jet. Fluctuations grow as they propagate through the disk and into the jet \citep[e.g.,][]{rieger_2019}, reaching peak modulation when they fully transfer energy to the jet base. If disk instabilities occur periodically, they can generate QPOs that naturally follow the observed amplitude evolution \citep[][]{czerny_accretion_disk}. 

\subsection{Jet-Origin Scenarios}
QPOS can also originate from processes intrinsic to the jet itself. In the scenario proposed by \citet[][]{wiita_cells_2011}, QPOs can arise from non-axisymmetric structures in the jet, most likely in the magnetic field. Because of this, when a shock wave moves outward through the jet, it interacts with these irregularities, causing variations in the emitted light. When the shock moves through a helical structure within the jet, the effect is similar to a change in the jet’s direction, resulting in an apparent shift in the viewing angle. If the jet is cylindrical and the shock velocity remains constant, the recurring boosts in flux would generate a periodic signal. In this case, the QPO persists if the shock traverses the helical structure and retains the same period if another similar event occurs. Additionally, \citet[][]{wiita_cells_2011} suggested that recurring boosts in the turbulent flow behind a propagating shock can also produce QPOs. If a dominant turbulent cell is present in the jet, Doppler boosting should be observed, leading to a QPO. However, as the turbulent cell gradually decays, the QPO amplitude will diminish over time due to the intrinsic stochastic nature of turbulence. Furthermore, the period of the QPO may vary since the size and/or velocity of the turbulent cell can also change. 

\subsubsection{Magnetohydrodynamic Instabilities}

Another scenario could involve magnetohydrodynamic (MHD) instabilities that can induce helical distortions in the jet, leading to periodic emission variations \citep[][]{dong_kink_2020}. The instability starts weakly, grows as the magnetic field lines twist further, reaches a peak when it is fully developed, and then declines as the field structure dissipates. MHD instabilities can produce magnetic reconnections, leading to the formation of plasma blobs within the jet, also known as plasmoids \citep[][]{plasmoids_sinori_2016}. During reconnection events, the formation and interaction of plasmoids can lead to rapid and intense $\gamma$-ray flares. As these plasmoids merge or are ejected from the reconnection site, they can produce variability patterns in the jet emission \citep[][]{magnetic_reconnections_plasmoids_shukla_2020}, associated with rapid (duration of hours to days) flare activities \citep[][]{meyer_sizeblobs_2021} or variations on timescales of months to years \citep[][]{timescale_plasmoids_christie_2019}. 

In addition to that, the plasmoids have also been proposed to be the origin of QPOs observed in the $\gamma$-ray emission of blazars. Specifically, in the model proposed by \citet{mohan_mangalam_blobs}, a plasmoid follows a helical path determined by the magnetic field geometry. Specifically, the plasmoid experiences a combination of toroidal and poloidal magnetic field components, leading to a spiral trajectory. The pitch angle of the helix, which governs the tightness of the spiral, is dependent on the relative strength of these magnetic components \citep[][]{asada_pitch_angle}. Moreover, the size of the plasmoid plays a role in the emitted flux. Larger plasmoids have longer light-crossing times, leading to extended flare durations and more symmetric LCs \citep[][]{finke_size_blobs_2024}. In contrast, smaller plasmoids produce shorter, sharper flares, which can result in more rapid variability in the LCs \citep[][]{meyer_sizeblobs_2021}. Additionally, the speed of the plasmoid within the jet significantly impacts the observed flux. Faster-moving plasmoids experience stronger Doppler boosting, leading to an enhancement in the observed intensity, whereas slower plasmoids contribute less prominently to the variability pattern \citep[][]{meyer_sizeblobs_2021}. Additionally, energy dissipation and flux decay are tied to plasmoid dynamics. Flux decay can be influenced by the radiative cooling of particles within the plasmoid and its expansion, which reduces its magnetic field strength and particle density, leading to a decrease in emission intensity over time \citep[][]{timescale_plasmoids_christie_2019}. 

As the plasmoid moves through the jet, it may encounter discontinuities in physical properties, such as velocity shear or heterogeneous-density regions. These interactions can generate shock waves, which, in turn, compress the plasma and enhance particle acceleration \citep[][]{giannios_blobs_envelope_2013}. This process leads to bursts of synchrotron and inverse Compton emission \citep[][]{shock_marscher_1985, shock_heavens_1988}. As a result, the amplitude of oscillations in the emission flux can increase due to the shocks amplifying the density and energy of pre-existing fluctuations \citep[][]{fichet_increase_trends_shock_2022}. The oscillation amplitude peaks when the shock achieves maximum efficiency in energizing particles, gradually declining as energy dissipates and the shock weakens. This energy dissipation could result in a progressive decrease in oscillation amplitude, marking the transition from an active acceleration phase to a more quiescent state \citep[][]{fichet_increase_trends_shock_2022}. 

\subsubsection{External Jet Agents}
Shock waves can also be generated by the interaction of external agents with the jet. One example is the penetration of dense clouds from the broad-line region into the jet. As these clouds interact with the relativistic flow, they induce strong shocks that efficiently accelerate particles, leading to enhanced synchrotron and inverse Compton emission, particularly in the $\gamma$-ray band \citep[][]{show_wave_clouds_palacio_2019}. 
Another source of shocks can be the collision between the jet and stellar winds from massive stars within the host galaxy. These winds inject additional turbulence and inhomogeneity into the jet, generating shocks that modify the particle acceleration processes and impact the emission \citep[][]{shock_wave_stellar_wind_bicknell_2002}. 

\subsubsection{Jet Precession}
Another phenomenon that can offer a compelling explanation for the dynamic evolution of the oscillation amplitude is the precession of the jet. In the scenario, changes in the Doppler boosting due to the variable orientation of the jet with respect to the line of sight cause changes in the amplitude of the oscillations over time \citep[][]{camenzind_jet}. This is particularly evident when the precessing jet crosses critical angles that either amplify or attenuate the emitted radiation, creating a distinct trend in the LC \citep[e.g.,][]{precession_chen_2024}. The precession could be originated by multiple scenarios. One scenario could be the Lense-Thirring precession \citep[e.g.,][]{franchini_lense, zanazzi_lense}. This phenomenon is a relativistic effect caused by the dragging of spacetime due to the rotation of an SMBH. In blazars, the jet is typically aligned with the accretion disk’s angular momentum axis. However, if the accretion disk is misaligned with the SMBH spin axis, the frame-dragging effect from the rotating SMBH induces precession in the inner regions of the disk. This Lense-Thirring precession gradually changes the orientation of the jet, leading to observable variations in the direction of the jet over time. Alternatively, the presence of a binary SMBH system at the center of the blazar can be the origin of such precession \citep[e.g.,][]{graham_binary, qian_precesion_binary}. In the binary scenario, gravitational torques within the SMBH system induce precession, causing the relativistic jet of one SMBH to follow a slow, precessional path influenced by its companion. This gradual angular shift changes the jet angle relative to our line of sight, modifying the Doppler boosting effect \citep[e.g.,][]{villata_helical_jet}. 

\subsection{Interpretation}\label{sec:interpretation}
In this paper, we interpret the observed QPOs with multiplicative amplitudes as the result of a combination of different phenomena. First, we consider that QPOs arise from a plasmoid following a helical trajectory dictated by the magnetic field geometry. As discussed in $\S$\ref{sec:theoretical_model}, this motion alters the angle of our line of sight with respect to the jet, leading to periodic variations in the Doppler boosting factor. Additionally, the movement of the plasmoid can generate shock waves, which, in turn, induce fluctuations in the emitted flux. This effect can be perceived as an apparent change in the jet’s axis and, consequently, in our viewing angle \citep[][]{wiita_cells_2011}, modulating the overall flux observed from the jet.   

Combined with this, we also use the conclusions presented in \citet{rieger_2004} to explore how differential Doppler boosting, caused by the helical motion of emitting components within the jet, can lead to observable periodicity. This motion is nonballistic, meaning the radiating plasma elements follow curved trajectories rather than moving in straight lines. \citet{rieger_2004} identified three possible physical mechanisms for helical jet motion in blazars. The first is nonballistic motion driven by the orbital motion in an SMBH binary. The second involves either ballistic or nonballistic helical jet paths induced by jet precession, which may result from gravitational torques exerted by a secondary black hole or from accretion disk warping. The third mechanism attributes nonballistic helical motion to intrinsic jet instabilities. The first two scenarios are particularly viable for explaining $P_{obs} \gtrsim$1 year. 

\subsubsection{Model expressions} \label{sec:equations_model}
The relativistic Doppler factor $\delta$ can be defined as
\begin{equation}
\label{eq:delta}
\delta=\frac{1}{\Gamma[1-\beta\cos(\theta)]},
\end{equation}
with $\Gamma$ being the Lorentz factor, $\beta$ the velocity of the jet in units of the speed of light, and $\theta$ the viewing angle. The presence of moving plasmoid combined with the presence of shock waves produces variations of the Doppler factor $\delta$ as a consequence of changes in the $\theta$ over the time $\theta_{obs}(t)$. We assume that $\theta_{obs}(t)$ changes according to the measured QPO period $P_{obs}$. In this scenario, the angle between the jet and the direction of the emitting region (pitch angle), $\phi$, and the angle between the jet and the line of sight, $\psi$, can be estimated as

\begin{equation}
\label{eq:cos_theta_time}
\cos\theta_{obs}(t) =\cos(\phi)\cos(\psi)+\sin(\phi)\sin(\psi)\cos(2\pi t/P_{obs}),
\end{equation}
as presented in \citet{otero_mwl}. Additionally, the emitted flux can be expressed as 
\begin{equation}
    \label{eq:flux}
    {F_{\nu} \propto F_{\nu^{'}}^{'}}{\delta^{-(n+\alpha)}},
\end{equation}
where $F_{\nu}$ and $F_{\nu^{'}}^{'}$ are the observed and rest-frame emission, and $n$ and $\alpha$ are the Doppler boosting factor and intrinsic spectral index, respectively. Combining Equation (\ref{eq:flux}) with Equations (\ref{eq:delta}) and  (\ref{eq:cos_theta_time}), the expression of the flux results in  

\begin{equation}
    \label{eq:flux_expression}
    \begin{split}
        F_{\nu} \propto  \frac{F_{\nu^{'}}^{'}}{\Gamma^{(n+\alpha)} (1 + \sin(\phi) \sin(\psi))^{(n+\alpha)}} \times \\
        \left[ 1 - \frac{\beta \cos(\phi) \cos(\psi)}{1 + \sin(\phi) \sin(\psi)} 
        \cos\left(\frac{2\pi t}{P_{obs}}\right) \right]^{-(n+\alpha)},
    \end{split}
\end{equation}

In this study, we approximate the emission model to capture the dynamic evolution of the plasmoid position within the jet over time, shaped by a combination of its shifting location due to its helical motion along the jet and the shock wave influence. The movement of the plasmoid is linked to the transient QPOs observed in the LC, which reflect its periodic motion within the jet. The shock wave adds layers of complexity to the plasmoid’s path, modulating the QPO, reflected as a linear temporal evolution of the amplitudes.

The combination of the plasmoid movement and the shock wave results in a continually shifting emission profile, as the orientation of the plasmoid and position relative to an observer change over time. This evolving geometry influences the observed emission of $\gamma$ rays, introducing specific temporal variability. 
As described by Equation~(\ref{eq:flux_expression}), in this emission model, the flux variations are solely governed by the dynamic evolution of the plasmoid, characterized by $\psi$ = $\psi(t)$.

\subsubsection {Flux Fit}\label{sec:amplitude_fit}
Examining the shape of the oscillations (see Figure \ref{fig:qpos_lcs}), we observe that their amplitude initially increases over time, reaching a peak, and subsequently decreases. 
Consequently, we substitute $\psi$ in Equation~(\ref{eq:flux_expression}) with a linear function as $\psi(t)$=$\pm at$. This approach suggests that the amplitudes evolve over time through a multiplicative relationship \citep[][]{penil_2025}. Therefore, our model accounts for the modulation introduced by the impact of the shock wave on the plasmoid's emission.
To extract the linear component of the flux, we employ a linear regression\footnote{We use the \texttt{LinearRegression} function of the \textsc{Python} package \texttt{Scikit-learn}, which is optimized specifically for linear fitting.}, characterizing the parameters of the linear trend\footnote{We use the \texttt{seasonal\_decompose} function of the \textsc{Python} package \texttt{Statsmodels}.}. 

Additionally, we perform a new QPO analysis to better characterize the oscillations. Specifically, we conduct a separate analysis of the QPOs for both the rising and falling segments of the LC using the methodology described in $\S$\ref{sec:methodology}. By evaluating whether each segment exhibits the same or distinct QPO characteristics, we aim to gain deeper insights into the flux behavior within the LC. This approach allows us to determine if the oscillatory patterns are consistent across different phases or if they vary, providing a more comprehensive characterization of the flux LC dynamics. 

We then test the hypothesis of the LC being accurately described by this model through the R$^{2}$ parameter, a statistical measure that assesses the goodness of fit of a regression model. The value of R$^{2}$ ranges from 0 to 1 (or 0 to 100 in percentage), where the higher values denote a better fit and suggest that the model can better explain the variations in the data. In \citet{hair_r2_2011}, three different levels of R$^{2}$—25\%, 50\%, and 75\%—are outlined and categorized as \textit{weak}, \textit{moderate}, and \textit{substantial}, respectively to evaluate the accuracy of a fit.

\begin{figure*}
	\centering
        \includegraphics[scale=0.165]{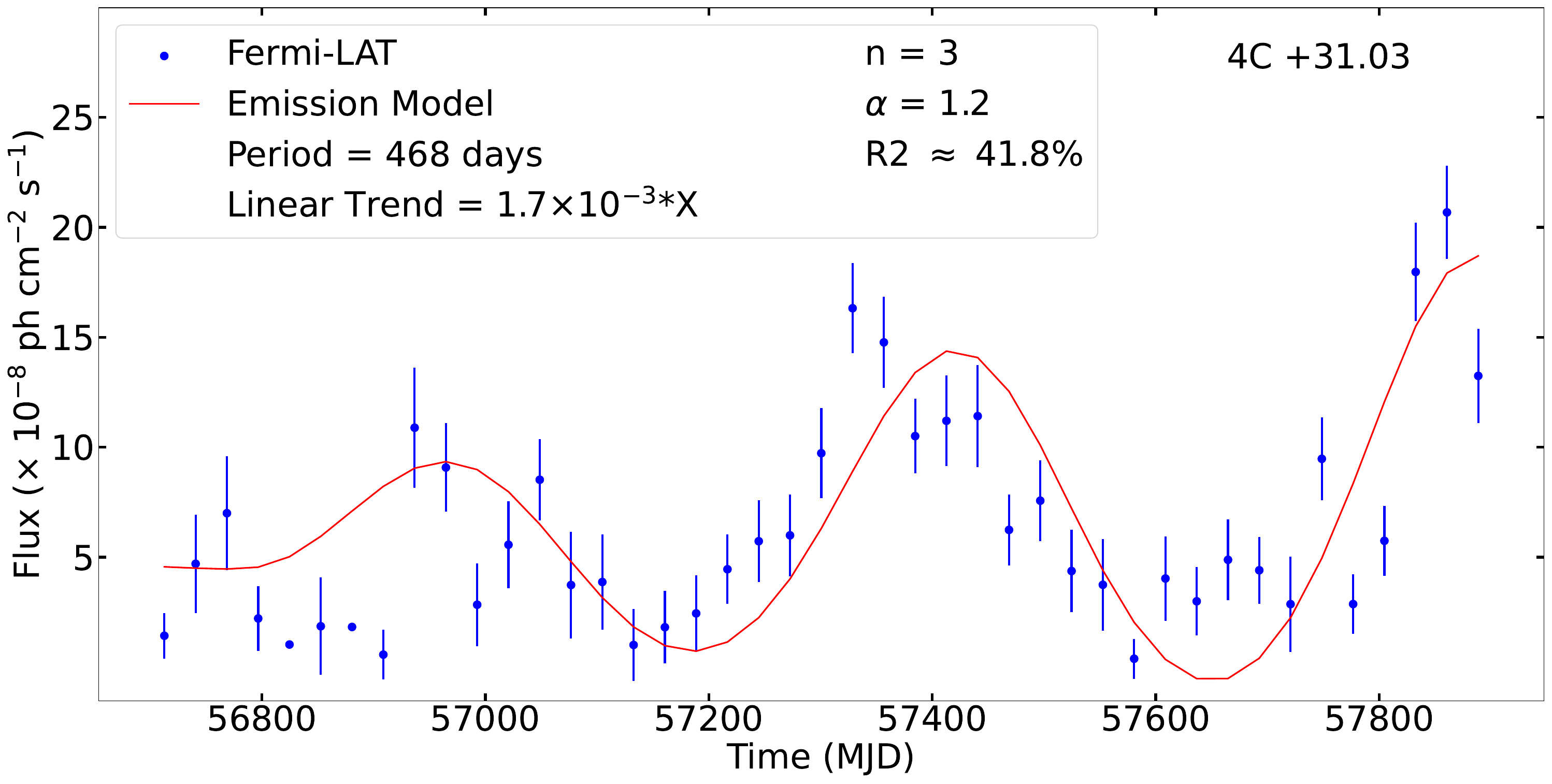}
	\includegraphics[scale=0.165]{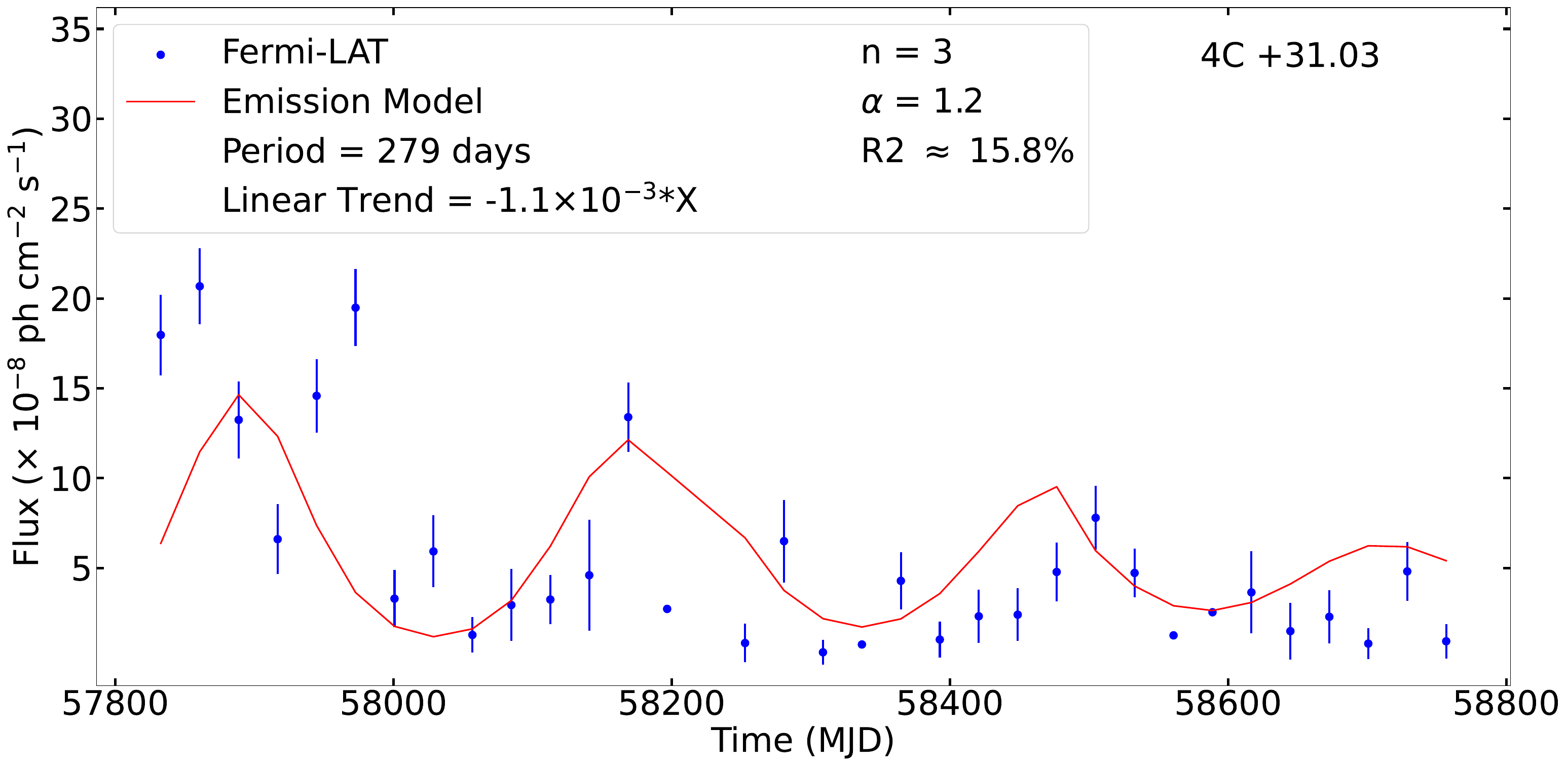}
        \includegraphics[scale=0.165]{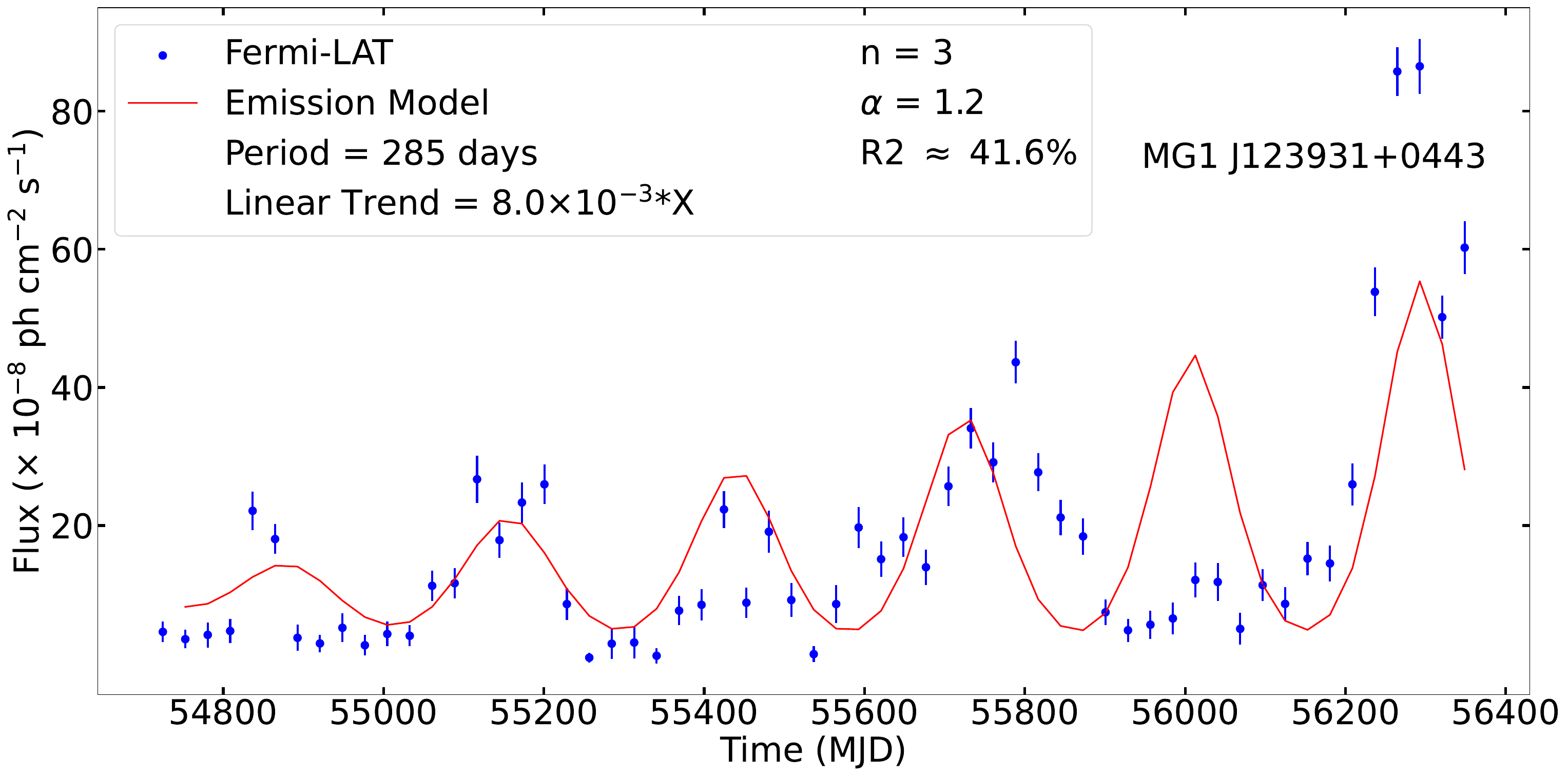}
        \includegraphics[scale=0.165]{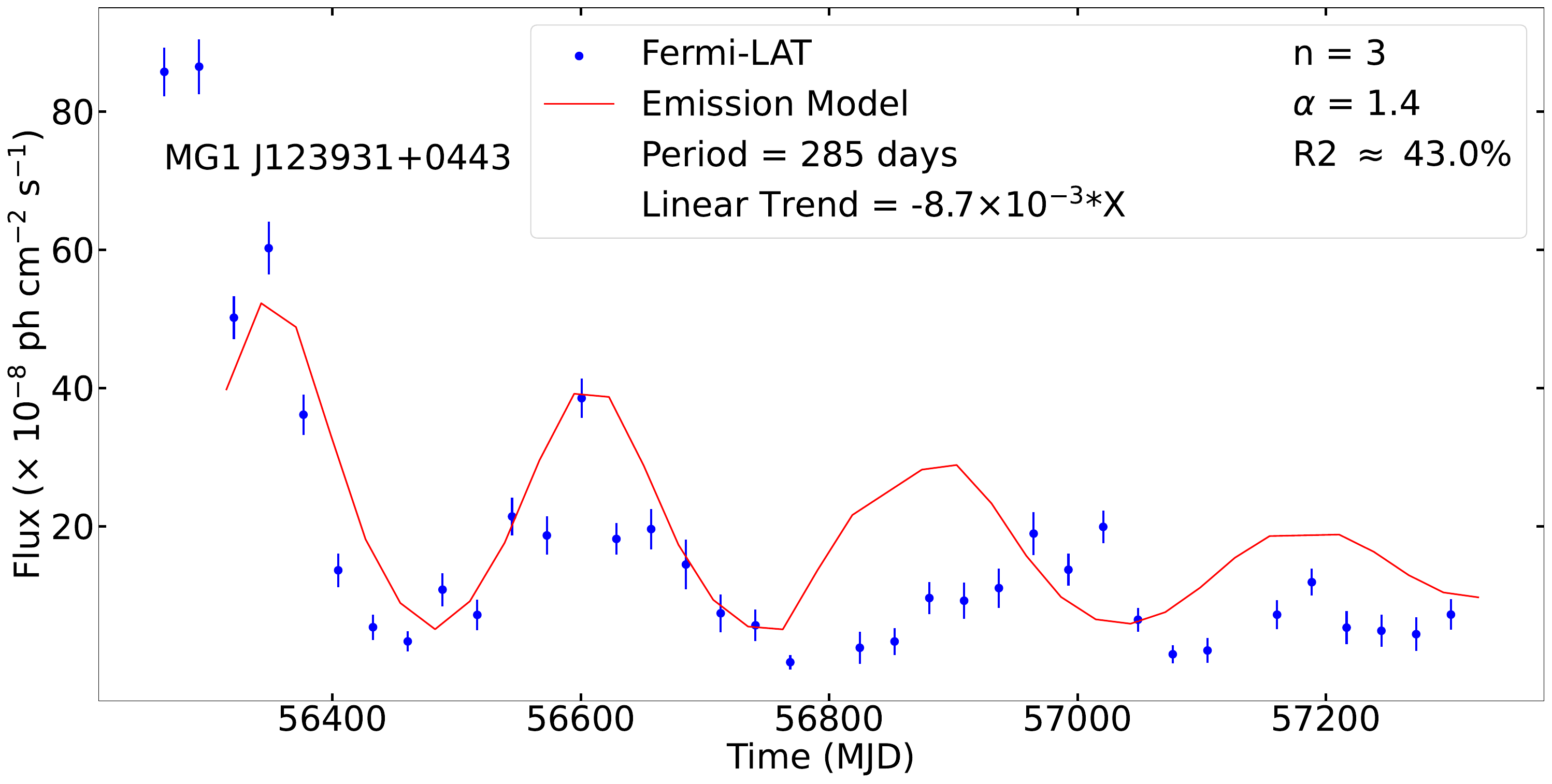}
        \includegraphics[scale=0.165]{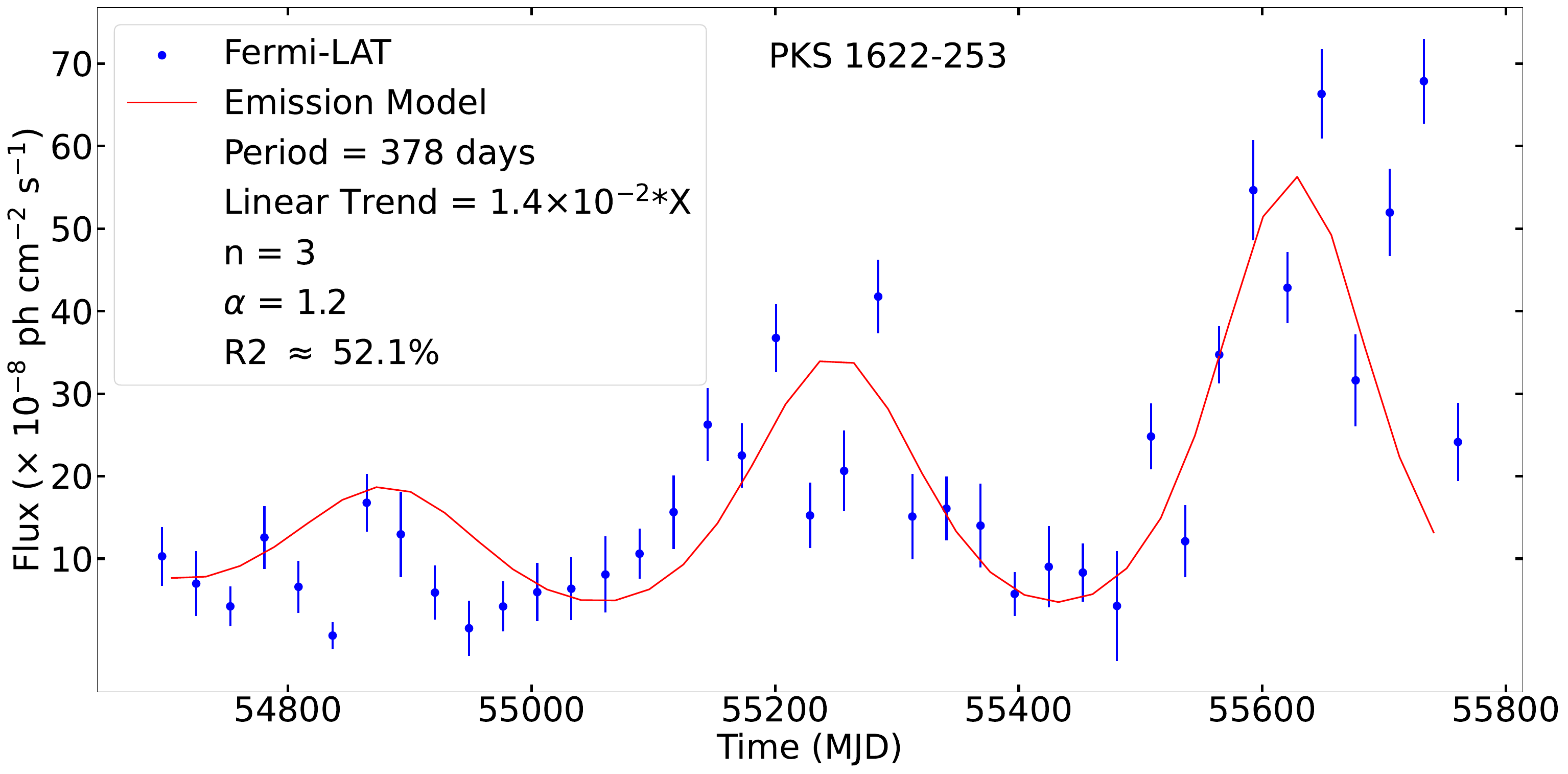}
        \includegraphics[scale=0.165]{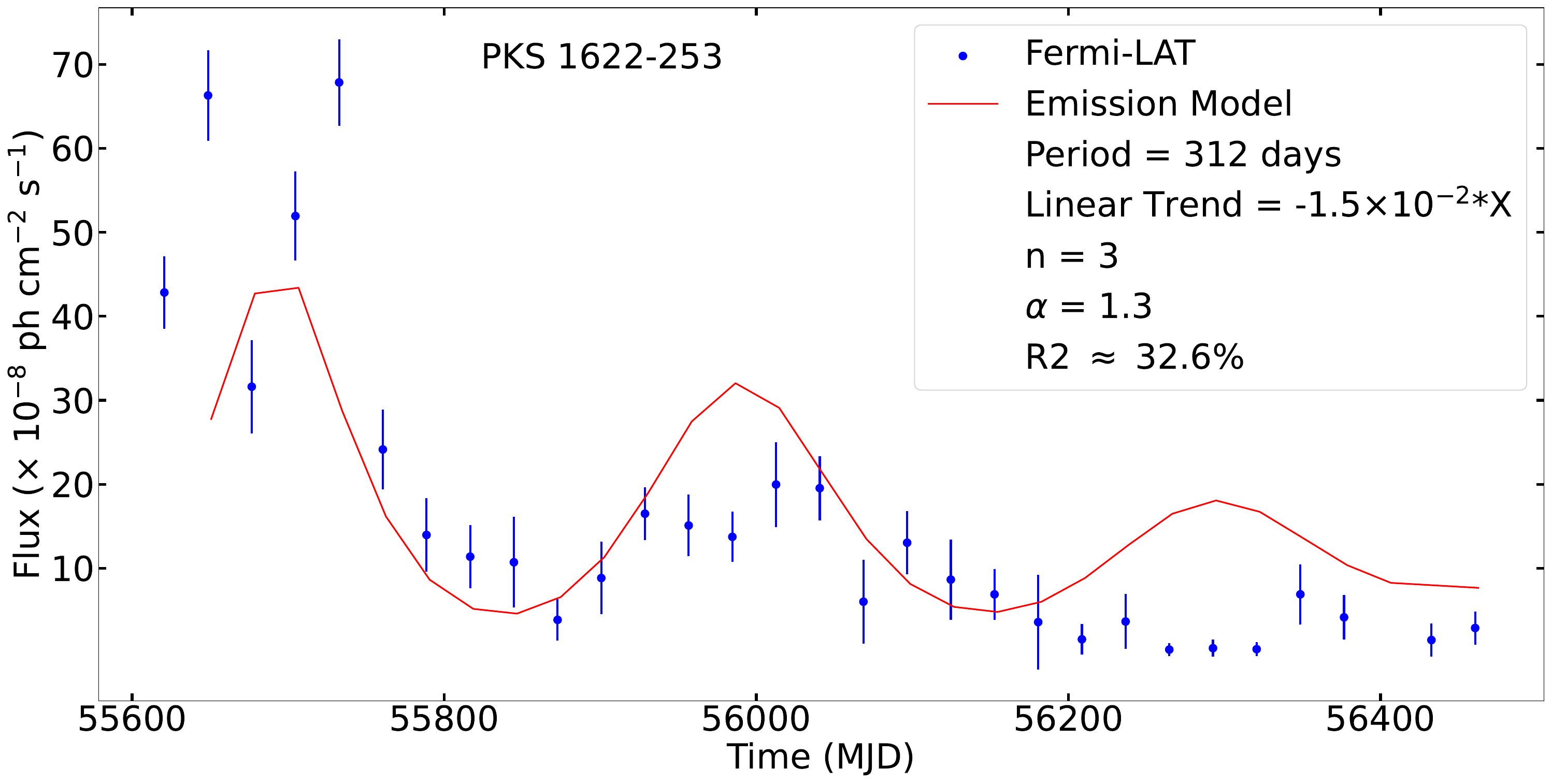}
	\caption{LCs from Figure \ref{fig:qpos_lcs}, illustrating individual segments with distinct flux trends. \textit{Left:} Increasing trends. \textit{Right:} Decreasing trends. \textit{From top to bottom:} 4C +31.03, MG1 J123931+0443 and PKS 1622$-$253. The red lines indicate the emission fit based on Equation (\ref{eq:flux_expression}). The plot legends display the fit parameters, showing the derived period and linear trend estimated for each segment to optimize the R${^2}$ value. The period and trend values correspond to thoe reported in Table~\ref{tab:fitting_results}.} \label{fig:flux_fits}
\end{figure*}

\begin{table*}
\centering
\caption{Results of the estimated period and linear fit for each segment of the transient QPO in the analyzed blazars. The table also lists the optimized period and linear fit values, as well as the Doppler boosting factor $n$, that maximize $R^{2}$, based on the previously determined ranges for periods and linear fits, as well as the explored range of $n$ and $\alpha$.}\label{tab:fitting_results}
\resizebox{\textwidth}{!}{%
\begin{tabular}{ccccc|ccccc}
\hline
\multirow{2}{*}{Source Name} & \multirow{2}{*}{Segment} & \multicolumn{2}{c}{Period [days]} & \multirow{2}{*}{Linear fit} & \multicolumn{2}{c}{Optimized Fit} & \multirow{2}{*}{$n$} & \multirow{2}{*}{$\alpha$} & R$^2$  \\ 
    &        &      SSA     &  GLSP  &  &  Period [days]   &  Linear fit  & & & [\%]   \\ \hline
\multirow{2}{*}{4C +31.03} &    Increasing      &  525$\pm$73 (3.4$\sigma$) & 554$\pm$84 (1.2$\sigma$)   & (2.0$\pm$0.4)$\times10^{-3}$    & 468 & 1.7$\times10^{-3}$ & 3 & 1.2 & 41.8  \\ \cline{2-10} 
                           &   Decreasing   &  303$\pm$48 (2.0$\sigma$)  & 318$\pm$77 (0.5$\sigma$) & (-1.7$\pm$0.7)$\times10^{-3}$ & 279 & -1.1$\times 10^{-3}$ & 3 & 1.2 & 11.2  \\ \hline
\multirow{2}{*}{MG1 J123931+0443} &  Increasing   &  291$\pm$26 (2.0$\sigma$) & 296$\pm$24 (1.0$\sigma$) & (8.3$\pm$0.6)$\times10^{-3}$ & 285 & 8.0$\times10^{-3}$ & 3 & 1.2 & 40.2 \\ \cline{2-10} 
                                  &   Decreasing     & 328$\pm$38 (2.5$\sigma$)  &  346$\pm$49 (2.0$\sigma$) & (-8.8$\pm$0.4)$\times10^{-3}$ & 285 & -8.7$\times10^{-3}$ & 3 & 1.4 & 43.0 \\ \hline
\multirow{2}{*}{PKS 1622$-$253} &  Increasing   &  420$\pm$66 (2.2$\sigma$) & 441$\pm$74 (1.0$\sigma$) & (1.8$\pm$0.6)$\times10^{-3}$ & 378 & 1.4$\times10^{-3}$ & 3 & 1.2 & 52.1 \\ \cline{2-10} 
                                  &   Decreasing     & 352$\pm$51 (1.5$\sigma$)  &  358$\pm$59 (0.5$\sigma$) &  (-2.0$\pm$0.6)$\times10^{-3}$ & 312 & -1.5$\times10^{-3}$ & 3 & 1.3 & 32.6 \\ \hline
\end{tabular}
}
\end{table*}

\section {Results and Discussion}\label{sec:flux_analysis}
The fit of the oscillations depends on variables such as the QPO period, the slope of linear trends, the intrinsic spectral index ($\alpha$), and the Doppler boosting factor ($n$), as detailed in Equation (\ref{eq:flux_expression}).
Therefore, we need to define adequate values of these parameters. In particular, for $\Gamma$, $\beta$, and $\phi$, we explored the literature to identify specific values of these parameters for the blazars analyzed in this study. However, no available measurements were found. Consequently, we adopt generic values that have been commonly used in similar studies. 

The Lorentz factor, $\Gamma$, generally falls within the range of approximately 10 to 20 for FSRQs \citep[][]{gamma_fsrq_ghisellini}, often reaching values $\sim 20-30$ \citep{xiong2014}. In this study, we adopt $\Gamma = 15$ \citep{ghisellini_gamma, sarkar_curved_jet}. The parameter $\beta$ is directly related to $\Gamma$ through the expression $\Gamma = 1 / (1 - \beta^{-2})^{1/2}$, which yields $\beta = 0.99777$. The pitch angle, $\phi$, is another variable with a potentially broad range, typically between $1.5^{\circ}$ and $10.0^{\circ}$ \citep[e.g.,][]{jorstad_pitch_angle, butuzova_pitch_angle}. However, owing to the lack of observational constraints in the literature, we adopt a pitch angle value of $\phi = 2^{\circ}$ as done in previous studies \citep[][]{sarkar_curved_jet, anuvab_curve_jet}.

For the values of the period $P$, the linear slope, the intrinsic spectral index $\alpha$, and the Doppler boosting factor $n$, we conduct an exploratory process to achieve the optimal flux fit modeled by Equation (\ref{eq:flux_expression}) for each LC by maximizing the R$^{2}$ value. For the periods and slopes, this optimization spans the range of periods and slopes determined through the methodology outlined in $\S\ref{sec:methodology}$. The resulting values are summarized in Table \ref{tab:fitting_results}.   

Regarding  $\alpha$, \citet{spectral_index_fsrq_gamma} conducted a comprehensive analysis of blazars detected by \textit{Fermi}-LAT, reporting a mean photon spectral index of $\approx$1.4. Consistently, we explore values for $\alpha$ in the range of 1.2, 1.3, 1.4, 1.5, and 1.6 to better understand the variations in the spectral behavior.

Finally, for the Doppler boosting factor $n$, we explore different values, from $n=3$ to $n=5$. When the emission in the jet is primarily due to synchrotron radiation, $n=2$. This lower value reflects the isotropic nature of synchrotron emission in the rest frame of the jet, typically prominent in radio to optical wavelengths and causing moderate flux variability in these bands \citep[][]{urry_n}. For $\gamma$ rays, which are often produced via inverse Compton scattering, $n$ tends to be higher, typically 3 or 4 \citep[][]{dermer_doopler}. This increase occurs because inverse Compton scattering --- and especially external Compton scattering, where seed photons from outside the jet are scattered --- is more sensitive to the Doppler factor. This phenomenon is particularly dominant in FSRQs, which constitute the objects in our sample. Finally, $n=5$ is particularly relevant in scenarios where short-lived or transient events are observed, suggesting that a localized, highly energetic region within the jet contributes to extreme, rapid flux variations characteristic of the FSRQ \citep[][]{dermer_doopler}.
The specific values of the periods, slopes, $\alpha$, and $n$ that maximize the R$^{2}$ value are displayed in Figure \ref{fig:flux_fits} and Table \ref{tab:fitting_results}, highlighting the most accurate fits achieved for each case.

\subsection{4C +31.03}
As discussed in $\S$\ref{sec:amplitude_fit}, we perform a detailed analysis of the QPOs by separately examining the increasing and decreasing segments of the LC, characterizing potential variations in amplitude or period between the rising and falling phases. We apply Equation (\ref{eq:flux_expression}) to model each segment, optimizing the variable values to maximize R$^{2}$. 

The increasing segment is limited on MJD 57900 when the flare amplitude is highest (see Figure \ref{fig:qpos_lcs}). The values of the explored variables are presented in Table \ref{tab:fitting_results}, getting an R$^{2}=41.8$\%. The period obtained that maximizes R$^{2}$ is 468 days, the value for the Doppler boosting factor $n$ is 3, and the intrinsic spectral index $\alpha$ is 1.2. The flux is shown in Figure \ref{fig:flux_fits}. The differences in R$^{2}$ obtained using values of $n$ compared to the optimized value range from 25\% to 50\% for $n=3$ and from 7\% to 12\% for $n=5$. Similar differences are observed for the other analyzed blazars, indicating a consistent sensitivity of $R^2$ to the Doppler boosting factor. Regarding the intrinsic spectral index, the differences across various values range from 0.5\% to 16\%, highlighting the sensitivity of R$^{2}$ to this parameter. For the decreasing segment, the resulting fit quality of R$^{2}$ = 15.8\%, indicating a poor fit. 

To illustrate the evolution of $\psi(t)$, we present Figure \ref{fig:psi_evolution}, which depicts its temporal variation based on different parametrization results for the three blazars. This visualization shows how $\psi(t)$ evolves over time, helping to assess the theoretical scenarios considered for interpreting the observed flux variations. In the specific case of 4C~+31.03, we observe that the viewing angle evolution shows a discontinuity of $\approx$30\% between the increasing and decreasing segments of the LC. Both the low R$^{2}$ value and this discontinuity suggest potential inconsistencies in modeling the decreasing segment. The heightened variability within this portion of the LC could be impacting the analysis accuracy, as fluctuations in the $\gamma$-ray emission introduce complexities that may not align well with the current model assumptions. These results raise questions about the suitability of the selected theoretical model in capturing the transient QPO behavior of 4C~+31.03, suggesting that an alternative approach may be necessary to better understand the emission mechanisms in this blazar. One possible scenario is that recurring boosts in the turbulent flow behind a propagating shock generate the QPO. Due to the stochastic nature of turbulence, variations in its properties can cause fluctuations in both the QPO period and the emitted flux \citep[][]{wiita_cells_2011}, leading to changes in the observed period, as seen in the case of 4C +31.03 (see Table \ref{tab:fitting_results}).

\subsection{MG1 J123931+0443}
\begin{figure*}
	\centering
        \includegraphics[scale=0.223]{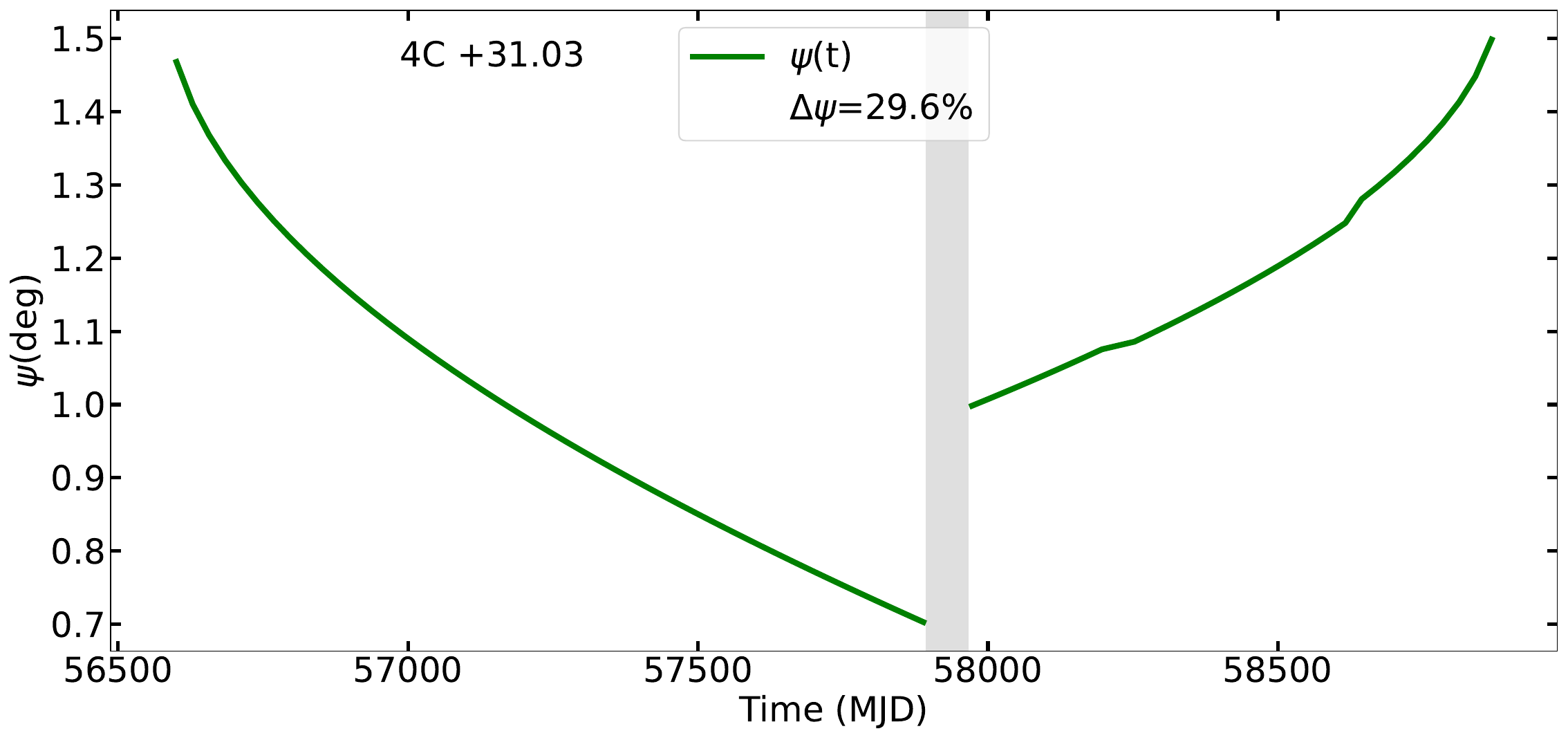}
	\includegraphics[scale=0.223]{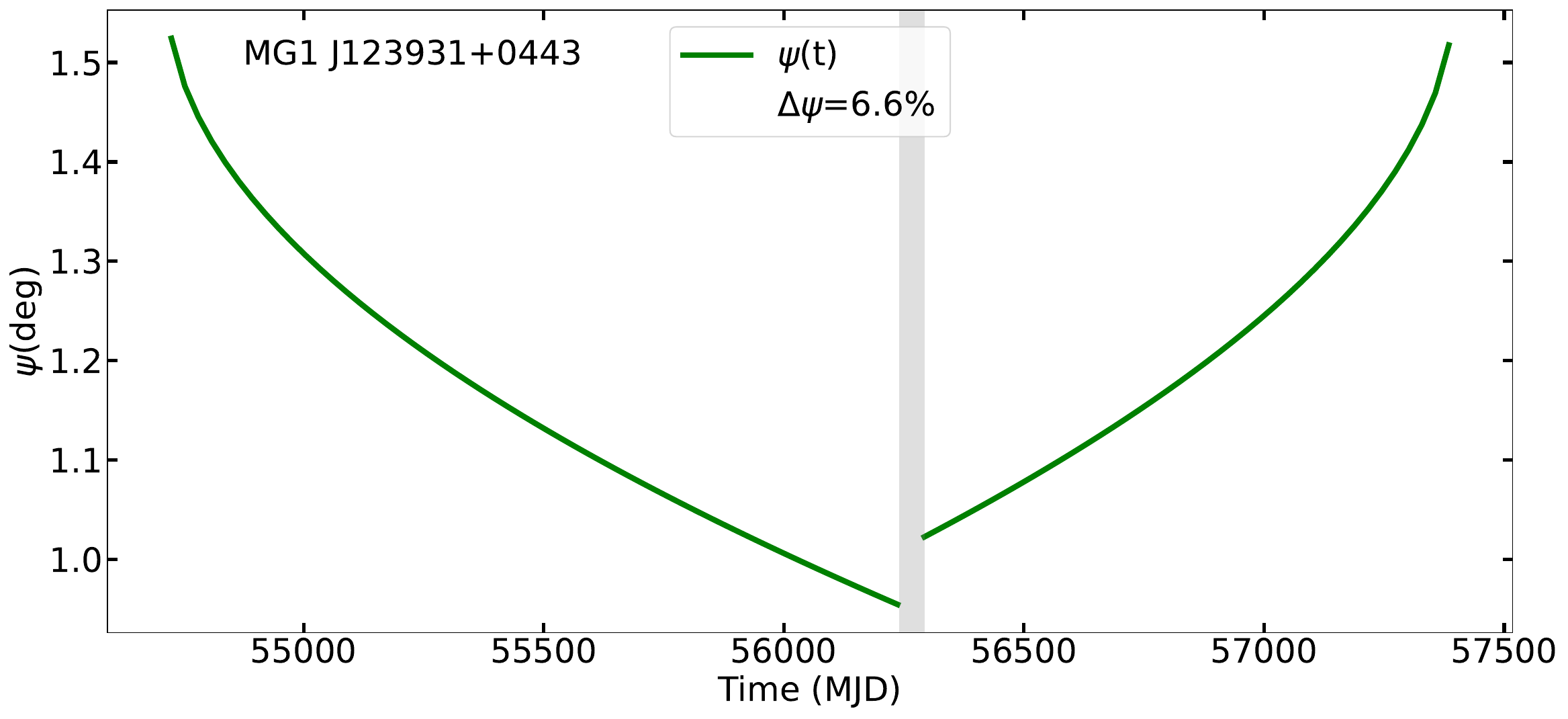}
        \includegraphics[scale=0.223]{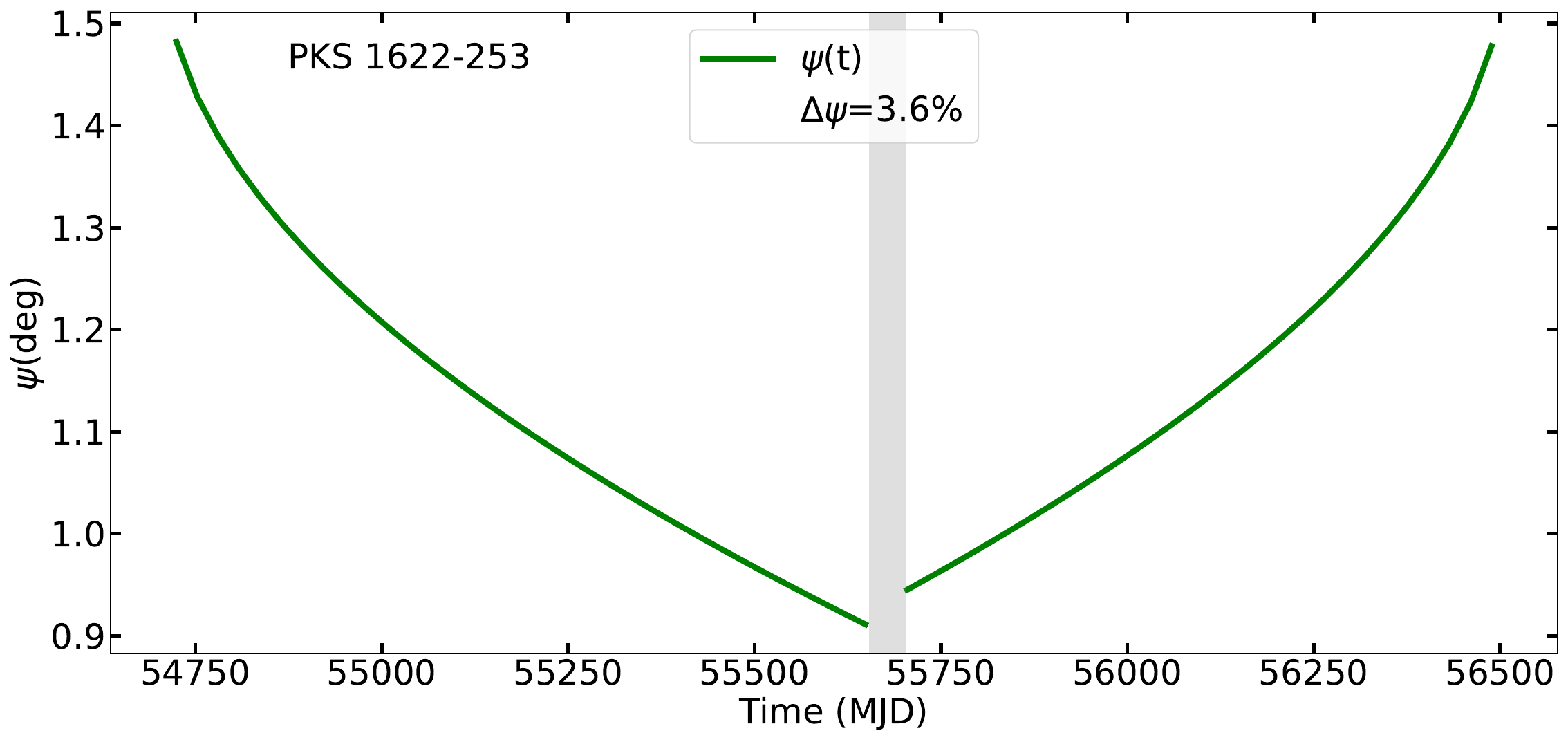}
	\caption{Jet viewing angle as a function of time. \textit{Top left}: 4C +31.03; \textit{Top right}: MG1 J123931+0443; \textit{Bottom}: PKS 1622$-$253. Grey vertical lines indicate the discontinuity arising from uncertainties in the trend estimation for each segment of the $\gamma$-ray emission. This separation reflects the differing observational segments, where variations in the estimated viewing angle may be attributed to fluctuations in the $\gamma$-ray emission trends. Such discontinuities underscore the sensitivity of the viewing angle to transient changes in jet orientation. The value $\Delta\psi$ quantifies this discontinuity between both segments of the LC.}
	\label{fig:psi_evolution}
\end{figure*}

The flux fits for both segments of the LC separated by the flare located at 56400 MJD present consistent results, with compatible periods and trends observed across both segments, as shown in Table~\ref{tab:fitting_results} and Figure \ref{fig:flux_fits}. Additionally, Figure \ref{fig:psi_evolution} reveals no significant discontinuity, with only a minor variation of 6.6\%, which can be attributed to uncertainties introduced by the flaring state during parameter estimation. In fact, from the QPO analysis, we observe that this error in the $\psi$ is consistent within errors at a 1$\sigma$ confidence level. These findings reinforce the accuracy of the chosen emission model, supporting it as a plausible scenario to explain the QPO observed in MG1 J123931+0443.

As discussed in $\S$\ref{sec:theoretical_model}, several theoretical scenarios can provide explanations for the observed QPO. \citet{rieger_2004} suggests that periodicity driven by differential Doppler boosting could result from (1) orbital motion in a close binary SMBH, (2) Newtonian precession within a close SMBH, or (3) internal jet rotation. Among these, Newtonian-driven precession appears particularly viable for explaining periods of $P_{obs} \approx 1$~year, which aligns with the observed period in MG1~J123931+0443, 285 days. 
If the period observed in this object is associated with the presence of a binary SMBH, we can relate $P_{obs}$ with the intrinsic orbital period ($P_{orbiting}$), as

\begin{equation}
\label{eq:period_orbiting}
P_{orbiting}=P_{obs}\frac{\Gamma^{2}}{1+z},
\end{equation}
where $z$ is the redshift of the source, and the motion of the plasma is considered as nonballistic and helical, driven by the binary system's orbital dynamics \citep[][]{rieger_2004}. A nonballistic motion occurs when the trajectory of the plasma is continuously influenced by external forces, such as the jet's environment or the central engine of the blazar \citep[][]{rieger_2004}. For an observed period $P_{obs}=285$ days and $z=1.761$ (see Table \ref{tab:candidates_list}), the calculated orbital period  $P_{orbiting}$ is $\approx64$~years assuming a $\Gamma=15$. 

Using this orbital period, we estimate the mass of the primary SMBH, $M_1$, based on the formulation provided by \citet{ostorero_2004},

\begin{equation}
\label{eq:mass_primary}
M_1\sim P_{orbiting}^{8/5} \left( \frac{M_1}{M_2} \right) ^{3/5}\times10^{6}M_{\odot},
\end{equation}
where the mass ratio $M_1/M_2$ typically ranges from $1$ to $100$ \citep{ostorero_2004}. Substituting the orbital period, we estimate the primary SMBH's mass to range from $\approx$7.7$\times10^{8}M_{\odot}$ to 1.2$\times10^{10}M_{\odot}$, consistent with reasonable mass estimates for SMBHs \citep[][]{liang_2003}. \citet{mg1_J123931_0443_massbh} estimated the SMBH mass of this blazar using various techniques, obtaining a range of 2.1-9.7$\times10^{8}M_{\odot}$, consistent with the range of values reported here. 

Using Kepler's Third Law, the orbital separation $a$ is determined as

\begin{equation}
\label{eq:orbital_separation}
a = \left( \frac{4\pi^2}{G(M_1 + M_2)} P_{\text{orbiting}}^2 \right)^{1/3}.
\end{equation}
$G$ is the gravitational constant and $M_{2}$ is the mass of the secondary SMBHs. This yields an orbital separation in the range of 8.9$\times10^{-2}$ to 1.8$\times10^{-1}$ parsecs. These values are consistent with those expected for binary SMBH systems, typically estimated to lie between $10^{-3}$ and $1$~pc \citep[][]{bogdanovic_2008, colpi_2014}.

If the primary SMBH emits a relativistic jet, its behavior may be influenced by the binary system's orbital dynamics. The orbital radius, defined as the distance between the primary SMBH and the system's barycenter \citep{ostorero_2004}, is calculated as:

\begin{equation}
\label{eq:orbital_radious}
r_1 = a \cdot \frac{M_2}{M_1 + M_2}.
\end{equation}

Using the estimated mass ranges, the orbital radius is found to range between 4.5$\times10^{-2}$ and 1.8$\times10^{-3}$ parsecs. 

\subsection{PKS 1622$-$253}
This blazar exhibits a high level of consistency between the estimated period and trend for both segments of the LC, further supported by the minimal discontinuity in $\psi$, measured at $\approx$3.6\%. The observed period, $P_{obs}\approx 340$ days, also suggests that Newtonian precession due to the interaction within a binary SMBH system could be a plausible physical scenario, as in the case of MG1 J123931+0443. 

We characterize the candidate binary SMBH system of PKS~1622$-$253. Using the redshift value of $z=0.786$ and the aforementioned expressions. We derive an orbital period $P_{orbiting}$ of $\approx$117 years. The estimation of the primary SMBH's mass ranges from $\approx$2.0$\times10^{9}M_{\odot}$ to 2.2$\times10^{10}M_{\odot}$, consistent with the mass of $\approx$4.0$\times10^{9}M_{\odot}$ reported by \citet{gamma_fsrq_ghisellini}. The  orbital separation ranges of 1.8$\times10^{-1}$ to 3.7$\times10^{-1}$ parsecs. The orbital radius ranges between 9.2$\times10^{-2}$ and 3.7$\times10^{-3}$ parsecs. 

\subsection{Precession Model}
An alternative scenario could be the precessing jet model, which was mentioned in $\S$\ref{sec:theoretical_model}. In this scenario, the relativistic jet emitted by a blazar undergoes precession, causing the jet's orientation to change over time. This variation alters the angle between the jet and our line of sight, influencing the observed emission due to changes in the relativistic beaming effects, potentially leading to the observed periodicity in the blazar's LCs \citep[][]{camenzind_jet}. 

In this analysis, the observed short-term multiplicative oscillations (those periodic signals identified over relatively brief observational intervals) do not directly represent the intrinsic precession period of the jet itself. Instead, these short-term oscillations emerge as secondary phenomena induced by the jet's precession, for instance, through the formation of a plasmoid. The plasmoid's presence significantly amplifies and modulates the observed flux, thereby generating observable short-term variability.

The true precession period, reflecting the jet's long-term behavior, would typically manifest as sustained, long-duration trends rather than short-lived, transient fluctuations. Detecting this intrinsic precession period is highly challenging due to the intrinsic characteristics of the analyzed LCs in this paper. 
The complete LCs of the blazars considered in this study (see Figure \ref{fig:blazar_lcs}) illustrate this difficulty explicitly. Aside from the analyzed segments exhibiting QPOs, the rest of the emission profiles are predominantly stochastic, characterized by flux variations consistent with random emission processes, as in the cases of MG1 J123931+0443 and PKS 1622$-$253. 

This inherent stochastic nature severely limits the possibility of reliably detecting and analyzing the intrinsic precession period. The intrinsic emission of the blazar itself is generally too faint to observe any modulation that may directly result from the jet's precession. It is only when an additional phenomenon occurs, such as the development of a plasmoid significantly enhancing the emission of the jet, that the modulation associated with the jet's precession becomes observable and analyzable.

Consequently, although we cannot definitively rule out jet precession as the underlying mechanism driving the observed modulated QPOs, the inherent complexities and stochastic nature of the analyzed LCs currently impede a robust and conclusive evaluation of such a theoretical model.

\section{Summary} \label{sec:summary}
In this paper, we analyzed the $\gamma$-ray emission of three blazars: 4C~+31.03, MG1 J123931+0443, and PKS 1622$-$253. Our analysis focuses on transient QPOs observed in their LCs, characterized by periodic oscillations with multiplicative amplitudes that locally vary linearly over time. We characterized these oscillations by determining both the period and the linear trend associated with each segment of the LC. Using these parameters, we modeled the $\gamma$-ray emission based on Doppler boosting effects.

The QPO features detected in PKS~1622$-$253 and MG1~J123931+0443 show a combination of year-like periodicities and smooth linear amplitude modulation, which align with the expectations for differential Doppler boosting induced by the presence of a binary SMBH system. In contrast, 4C~+31.03 exhibits less coherence, more typical of stochastic variability. Therefore, PKS~1622$-$253 and MG1~J123931+0443 emerge as binary SMBH candidates. However, given the relatively low significance derived from our transient QPO analysis, these results should be interpreted cautiously as tentative hints rather than definitive evidence. Additional studies and observations will be essential to further test and validate this interpretation. Despite these limitations, the consistency between the theoretical model and the observed $\gamma$-ray LCs highlights the plausibility that binary SMBH systems could be involved in producing QPOs in blazars, offering potentially valuable insights into the nature of periodic variability in these sources.

\section*{Acknowledgements}
\begin{acknowledgements}
P.P. and M.A. acknowledge funding under NASA contract 80NSSC20K1562. J.O.-S. acknowledges founding from the Istituto Nazionale di Fisica Nucleare (INFN) Cap. U.1.01.01.01.009. 

This work was supported by the European Research Council, ERC Starting grant \textit{MessMapp}, S.B. Principal Investigator, under contract no. 949555, and by the German Science Foundation DFG, research grant “Relativistic Jets in Active Galaxies” (FOR 5195, grant No. 443220636).
\end{acknowledgements}

\bibliographystyle{mnras}
\bibliography{oja_template}

\clearpage
\appendix
\noindent
\section{Complete $\gamma$-ray LCs}\label{sec:appendix}
\renewcommand{\thefigure}{A\arabic{figure}}
\setcounter{figure}{0}
This section includes the complete $\gamma$-ray LCs of the blazars studied in this paper, shown in Figure \ref{fig:blazar_lcs}. 
\begin{figure*}[htp!]
	\centering
        \includegraphics[scale=0.223]{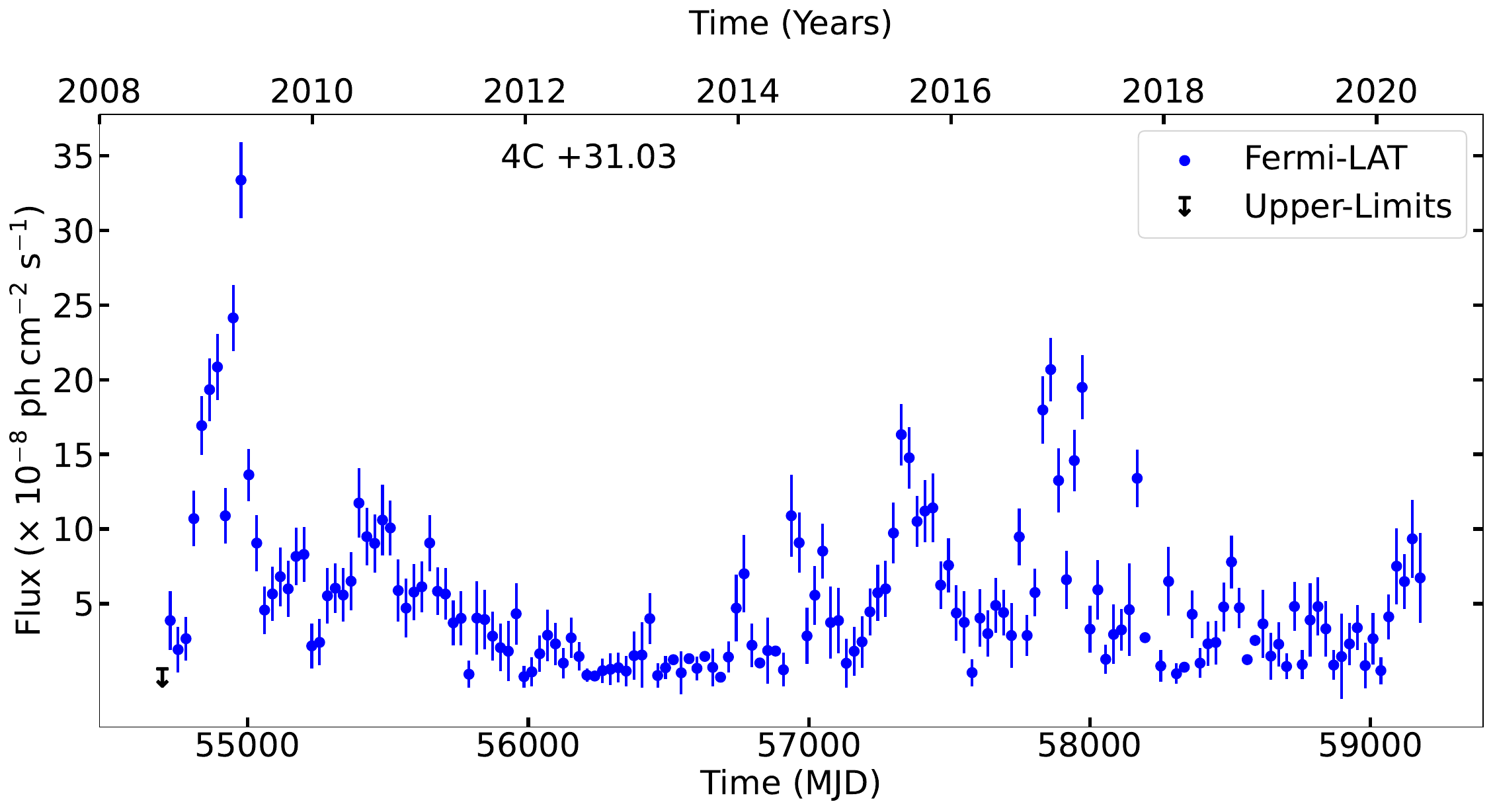}
	\includegraphics[scale=0.223]{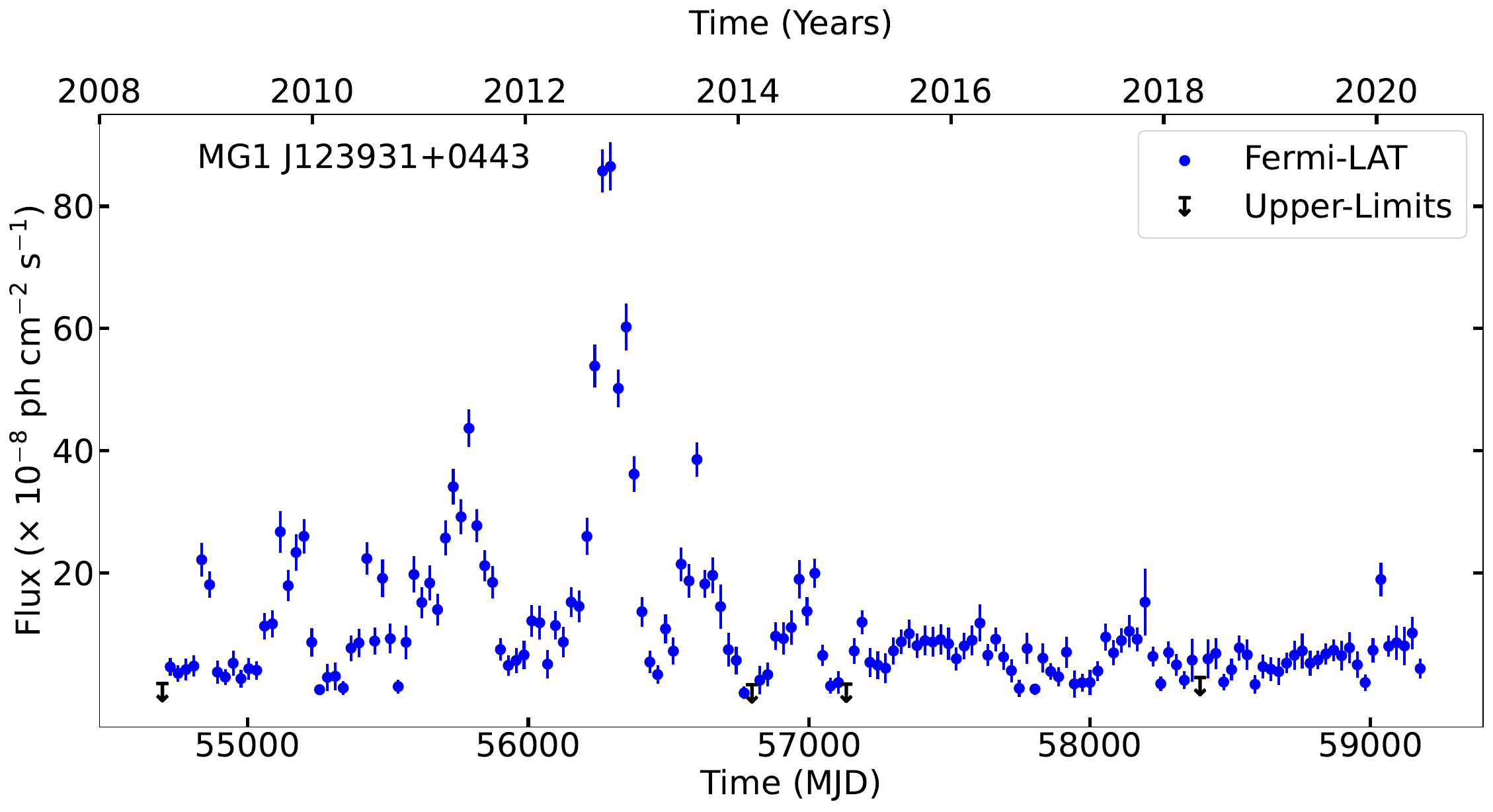}
        \includegraphics[scale=0.223]{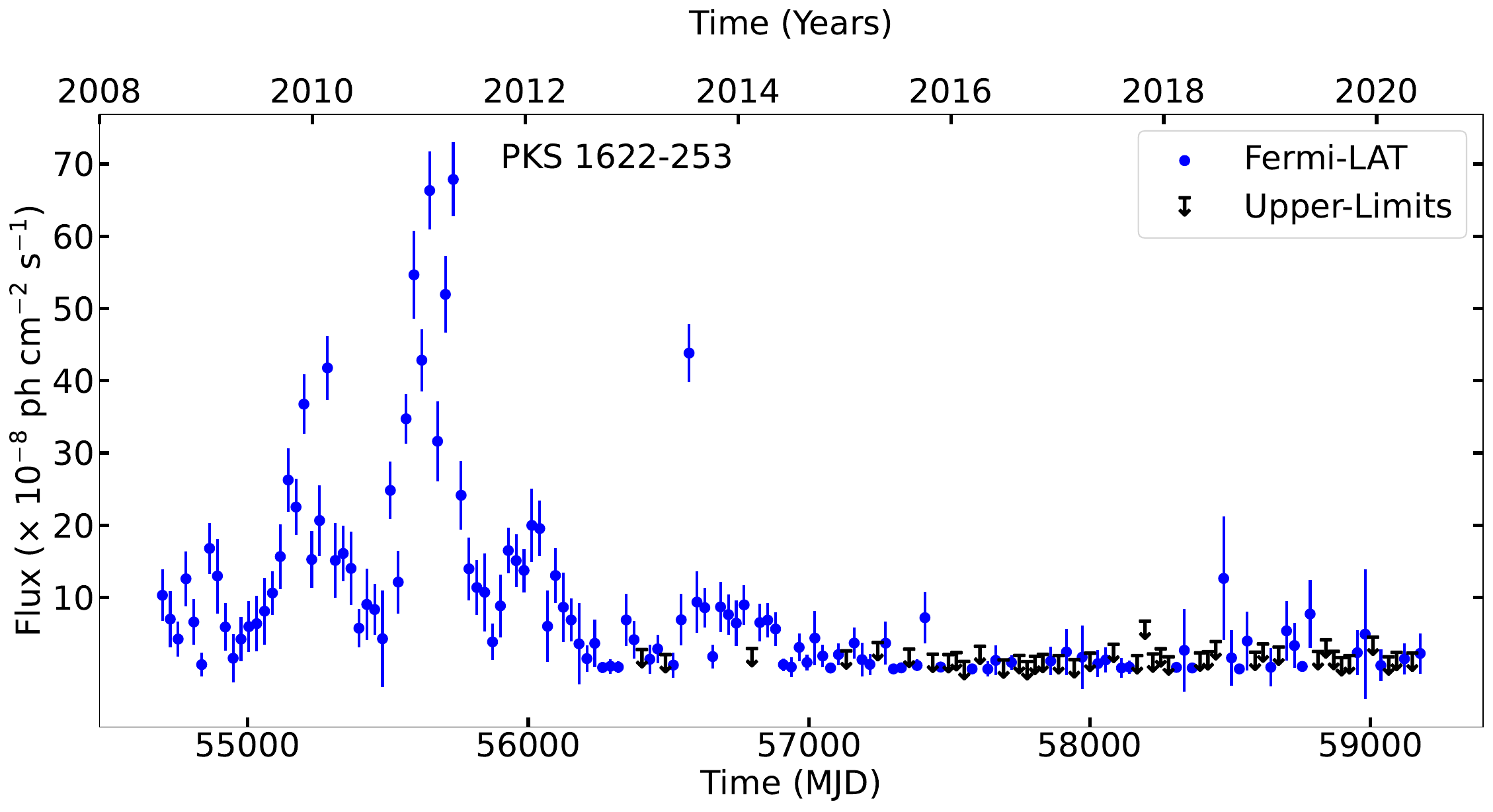}
	\caption{LCs of the blazars presented in Table \ref{tab:candidates_list}. \textit{Top left}: 4C +31.03. \textit{Top right}: MG1 J123931+0443. \textit{Bottom}: PKS 1622$-$253.} \label{fig:blazar_lcs}
\end{figure*}

\end{document}